\documentclass{article}%
\usepackage{amsmath}
\usepackage{amsfonts}
\usepackage{amssymb}
\usepackage{graphicx}%
\newtheorem{theorem}{Theorem}

\newtheorem{definition}[theorem]{Definition}

\newtheorem{remark}[theorem]{Remark}

\newcommand{\bpartial}{\mathop{\partial\kern -4pt\raisebox{.8pt}{$|$}}}
\newcommand{\bra}{\mathopen{[\kern-1.6pt[}}
\newcommand{\ket}{\mathclose{]\kern-1.5pt]}}
\newcommand{\bbra}{\mathopen{[\kern-2.2pt[\kern-2.3pt[}}
\newcommand{\bket}{\mathclose{]\kern-2.1pt]\kern-2.3pt]}}

\newcommand{\slg}{\mbox{\bfseries\slshape g}}
\newcommand{\sslg}{\mbox{\tiny \bfseries\slshape g}}

\makeindex
\begin{document}

\title{Pair and Impair, Even and Odd Form Fields, and Electromagnetism}
\author{Rold\~{a}o da Rocha$^{(1)}$ and Waldyr A. Rodrigues Jr.$^{(2)}$\\$^{(1)}\hspace{-0.05cm}${\footnotesize Centro de Matem\'atica, Computa\c c\~ao
e Cogni\c c\~ao}\\{\footnotesize Universidade Federal do ABC, 09210-170, Santo Andr\'e, SP,
Brazil}\\{\small \texttt{roldao.rocha@ufabc.edu.br}}\\$^{(2)}\hspace{-0.05cm}${\footnotesize Institute of Mathematics, Statistics
and Scientific Computation}\\{\footnotesize IMECC-UNICAMP CP 6065}\\{\footnotesize 13083-859 Campinas, SP, Brazil}\\{\small \texttt{walrod@ime.unicamp.br \textrm{or} walrod@mpc.com.br}}}
\date{}
\maketitle

\begin{abstract}
In this paper after reviewing the Schouten and de Rham definition of
\textit{impair} and \textit{pair} differential form fields (not to be confused
with differential form fields of even and odd grades) we prove that in a
\textit{relativistic spacetime} it is possible (despite claims in contrary) to
coherently formulate electromagnetism (and we believe any other physical
theory) using only pair form fields or, if one wishes, using pair and impair
form fields together, in an appropriate way\textit{.} Those two distinct
descriptions involve only a mathematical choice and do not seem to lead to any
observable physical consequence if due care is taken. Moreover, we show in
details that a formulation of electromagnetic theory in the Clifford bundle
formalism of differential forms where the two Maxwell equations of the so
called free metric approach becomes a single equation is compatible with both
formulations of electromagnetism just mentioned above. In addition we derive
directly from Maxwell equation the density of force (coupling of the
electromagnetic field with the charge current) that is a postulate in the free
metric approach to electromagnetism. We recall also a formulation of the
engineering version of Maxwell equations using electric and magnetic fields as
objects of the same nature, i.e., without using polar and axial vectors.

\end{abstract}

\section{Introduction}

Cartan has popularized the use of differential forms which he apparently
introduced in 1899 \cite{cartan00}, and which are now indispensable tools in
several mathematical and physical theories\footnote{In particular, it seems
that Cartan applied differential forms in the formulation of electromagnetism
for the first time in \cite{cartan0}.}. What is less known among physicists is
that those objects come out in two versions, \textit{pair} and \textit{impair}
differential forms (also called by some authors pseudo-forms or twisted
forms)\footnote{Impair forms are also called by some authors pseudo-forms or
twisted forms.}, a concept which has its origin in the Heaviside formulation
of electromagnetic theory in terms of \textit{polar} and \textit{axial} vector
fields. Rigorously speaking, pair and impair forms are sections of different
bundles\footnote{Pair forms are sections of the exterior algebra bundle $%
%TCIMACRO{\dbigwedge }%
%BeginExpansion
{\displaystyle\bigwedge}
%EndExpansion
T^{\ast}M$ and impair forms are sections the bundle $%
%TCIMACRO{\dbigwedge \nolimits_{-}}%
%BeginExpansion
{\displaystyle\bigwedge\nolimits_{-}}
%EndExpansion
T^{\ast}M\simeq$ $L(M)\otimes%
%TCIMACRO{\dbigwedge }%
%BeginExpansion
{\displaystyle\bigwedge}
%EndExpansion
T^{\ast}M$ where $L(M)$ is a line bundle called the orientation bundle of $M$.
Some details are given below.}, but here to motivate our presentation we may
say that pair forms living on an oriented spacetime are invariant under change
of the coframe basis orientation (related to a fixed spacetime orientation) in
which they are expressed\footnote{The orientation of a given coframe basis is
not to be confused with the orientation of the manifold (part of the structure
defining a spacetime) which is given by an \textit{arbitrary} choice of a
volume form. See below for details.} --- in particular, pair $0$-forms are
\textit{scalar} functions --- whereas impair forms change sign under change of
the coframe basis orientation in which they are expressed, and in particular
\textit{impair} $0$-forms are also known as \textit{pseudoscalar} functions.

A definition\footnote{We will give an alternative equivalent definition
below.} of such pair and impair differential forms has been originally
introduced by de Rham \cite{derham} (but see also \cite{schouten,weyl}%
)\footnote{It must be said that that in de Rham's discussion of cohomology,
impair forms have disappeared. In \cite{nosso1}, this question is suitably
studied by the investigation of the Grassmann and Clifford algebras over Peano
spaces, introducing their respective associated extended algebras, and
exploring these concepts also from the counterspace viewpoint. It was shown
that the de Rham cochain, generated by the codifferential operator related to
the regressive product, is composed by a sequence of exterior algebra
homogeneous subspaces that are subsequently pair and impair.} and will be
recalled below.

Of course, the theory of differential forms has been applied by many authors
in the formulation of different physical theories (see, e.g.,
\cite{rodoliv2007}), and in particular in electromagnetism. However, the
formulations of that theory appearing, e.g.,
\cite{bt,choquet,darling,felsager,flanders,goschu,huvan,parrott,thirring} make
use only of pair differential forms\footnote{Some authors as e.g.,
\cite{delphenich} avoid the use of pair and impair forms by using instead
(pair) multivector fields and pair forms.}. It must be said that for those
authors, the arena where charged particles and the electromagnetic field
interact is a \textit{Lorentzian spacetime}, that as well known is an
\textit{oriented }manifold\footnote{Almost all the authors in
\cite{bt,choquet,darling,felsager,flanders,goschu,huvan,parrott,thirring} even
do not mention impair differential forms in their books and the few that
mention those objects only say that they are necessary for a consistent
integration theory on non oriented manifolds.}. On the other side authors
like, e.g., \cite{burke,deschamps,
delphenich,edelen,hehl,hehl1,itin,janc2,kiehn,post,post1,post2,punt}%
\ explicitly claim that impair forms are absolutely necessary for a consistent
formulation of electromagnetism even in an \textit{oriented} spacetime
manifold and mainly if the spacetime is a bare manifold devoid of metric and
affine structure\footnote{The idea of developing electromagnetism using a
manifold devoid of metric and affine structure is very old and appears in
\cite{dan} and \cite{ko}. A complete set of references on the subject up to
1960 is given in \ \cite{truto}. Such an approach to electromagnetism has also
been used in \cite{scho} and now is advocated by many authors, see specially
\cite{hehl} \ (and of course, the arXiv) for modern references.
\par
{}}. Eventually, the main \textit{argument} of the majority of those authors
is that the current $3$-form must be impair for otherwise its integral over an
oriented $3$-chain (which gives the value of the charge in that region) does
depend on the orientation chosen, a conclusion that those authors consider an absurd.

Moreover it must be said that the presentation of the differential equations
of electromagnetism using the Clifford bundle formalism \cite{rodoliv2007}
uses only pair differential forms and, if the charge argument is indeed
correct, it seems to imply that the Clifford bundle cannot be used to describe
electromagnetism or any other physical theory. So, we must discuss in a
thoughtful way the claims of \cite{burke,deschamps,
delphenich,edelen,hehl,hehl1,itin,janc2,kiehn,post,post1,post2,punt,tonti} and
indeed, the main purpose of the present paper is to do that by showing that in
a \textit{relativistic spacetime}\footnote{The concept of a relativistic
spacetime as used in this paper is recalled in Section 2 and Subsection 3.2.}
the electromagnetic theory\footnote{This includes even the case of regions
involving a non dispersive medium that can be described by effective
Lorentzian spacetimes \cite{post}.} can be rigorously presented with all
fields involved being pair form fields. Of course, a presentation of
electromagnetism in a oriented (even if bare) spacetime using appropriate pair
and impair form fields is also \textit{correct}, but as it will become clear
below it seems to be nothing more than a simply \textit{option}, not a
\textit{necessity}. Moreover, we show that contrary to a first expectation,
the formulation of electromagnetism in the Clifford bundle \cite{rodoliv2007}
of (\textit{pair}) form fields\ is automatically compatible with each one of
those mentioned formulations of the theory, i.e., starting from Maxwell
equations formulated as a single equation in the Clifford bundle, we can show
that from that equation we can either obtain as a result of a straightforward
mathematical\textit{ choice }two equations involving only pair forms
\textit{or} two equations such that one uses a pair form and the other impair forms.

This paper is organized as follows: in Section 2 we introduce the nature of
the spacetime manifold used in the formulation of relativistic physical
theories and recall Maxwell equations formulated with \textit{pair}
differential forms on Minkowski spacetime, calling the reader's attention to
the fact that Maxwell equations describe only one aspect of electromagnetism,
which is a theory describing the interaction of the electromagnetic field with
charged particles (see specially Section 6). Moreover, we emphasize that
although only the manifold structure of $M$ is enough for the writing of
Maxwell equations, the remaining objects which defines the Minkowski spacetime
structure play a fundamental role in the theory, as showed on several times in
different sections of the paper. Section 3 is dedicated to the definition of
the pair volume 4-form and the pair Hodge star operator. Section 4 defines
\textit{impair} differential forms and in particular emphasizes the difference
between the pair and the impair volume forms and the pair and impair Hodge
star operators. Section 4 also discusses the fundamentals of electromagnetism
in a medium and proves that, contrary to some claims \cite{hehl2}, the recent
discovery that the constitutive extensor of Cr$_{2}$O$_{3}$ has a term
proportional to the Levi-Civita symbol in no way implies that this discovery
is the proof that impair forms must be used in the formulation of
electromagnetism. Section 5 recalls the Clifford bundle formulation of Maxwell
equation\footnote{No misprint here. To know why look at Eq.(52).}, proving as
already mentioned that it is compatible with those two formulations (only pair
and pair and impair) of that equations. Section 6 \ shows how the force
density that is postulated in the presentation of electromagnetism in
\cite{hehl} is directly contained in Maxwell equation. Moreover we show in
Section 6 that the equation (which contains the force density) describing the
interaction of the charged particles with the field automatically knocks down
the \textit{charge argument} mentioned above. In Section 7 using the Pauli
algebra bundle we present the engineering formulation of electromagnetism in
terms of the electric and magnetic fields $\mathbf{E}$ and $\mathbf{B}$ and
emphasize that in this formulation which necessarily needs a \textit{choice}
of a volume element we do \textit{not} need to introduce the so called
\textit{axial} vector fields and moreover that the circulation of the magnetic
field around a (very long) wire conducting current is \textit{conventional}.
Finally in Section 8 we present our concluding remarks.

\section{Nature of the Spacetime Manifold and of the Electromagnetic Field.}

Every physical theory starts by modeling the arena (spacetime) where physical
phenomena are supposed to happen. It is a well known fact that when
gravitation can be neglected, the motion (classical or quantum) of particles
and fields occurs in an arena which is modeled by Minkowski spacetime, i.e., a
structure $(M,%
%TCIMACRO{\TeXButton{slg}{\slg}}%
%BeginExpansion
\slg
%EndExpansion
,D,\tau_{{%
%TCIMACRO{\TeXButton{slg}{\sslg}}%
%BeginExpansion
\sslg
%EndExpansion
}},\uparrow)$, where $M$ is a 4-dimensional manifold diffeomorphic to
$\mathbb{R}^{4}$, $%
%TCIMACRO{\TeXButton{slg}{\slg}}%
%BeginExpansion
\slg
%EndExpansion
\in\sec T_{0}^{2}M$ is a Lorentzian metric, $D$ is the Levi-Civita connection
of $%
%TCIMACRO{\TeXButton{slg}{\slg}}%
%BeginExpansion
\slg
%EndExpansion
$ (i.e., $\mathbf{T(}D\mathbf{)}=0$, where $\mathbf{T}$ is the torsion tensor
associated to the connection $D$), $\mathbf{R(}D)$\textbf{ }is the curvature
tensor associated with $D$, \ $\tau_{%
%TCIMACRO{\TeXButton{slg}{\sslg}}%
%BeginExpansion
\sslg
%EndExpansion
}\in\sec%
%TCIMACRO{\dbigwedge \nolimits^{4}}%
%BeginExpansion
{\displaystyle\bigwedge\nolimits^{4}}
%EndExpansion
T^{\ast}M$ is the metric volume element, i.e., a \textit{pair}\footnote{See
below for the definition of \textit{pair} and \textit{impair} forms.} $4$-form
defining a spacetime orientation and $\uparrow$ denotes time
orientation\footnote{More details may be found, e.g. in
\cite{rodoliv2007,sawu}.}.

Classical electromagnetism according to Feynman is the theory which describes
the interaction of objects called charged particles and the electromagnetic
field $F\in\sec%
%TCIMACRO{\dbigwedge \nolimits^{2}}%
%BeginExpansion
{\displaystyle\bigwedge\nolimits^{2}}
%EndExpansion
T^{\ast}M$ called field strength. For the purposes of this paper a charged
particle is described by a triple $(m,q,\sigma)$, where (conventionally)
$m\in\mathbb{R}^{+}$ is the \textit{mass} parameter and based on experimental
facts $q$ (the charge) is a non null integral multiple of an elementary charge
denoted $\left\vert e\right\vert $. It is extremely important to keep in mind
for the objectives of the present paper that the sign of $q$ to be attributed
to any charge depends on a \textit{convention}, which will be scrutinized
latter (Remark 17 and Section 4.3).\ Moreover, $\sigma:\mathbb{R\rightarrow
M}$ is timelike curve pointing to the future\footnote{This being the reason
why we suppose that spacetime is time orientable.}. We parametrize $\sigma$ in
such a way that $%
%TCIMACRO{\TeXButton{slg}{\slg}}%
%BeginExpansion
\slg
%EndExpansion
(\sigma_{\ast},\sigma_{\ast})=1$ and define a $1$-form field over $\sigma$,
denoted by $v=%
%TCIMACRO{\TeXButton{slg}{\slg}}%
%BeginExpansion
\slg
%EndExpansion
(\sigma_{\ast})$. Given a finite collection of particles $(m^{(i)}%
,q^{(i)},\sigma^{(i)})$, $i=1,2,\ldots,n$, we define the \textit{current} for
the $i$-particle as the $1$-form field $J^{(i)}=q^{(i)}$ $v^{(i)}$ over
$\sigma$. The total current of the system is given by%
\begin{equation}
J=%
%TCIMACRO{\dsum \nolimits_{i}}%
%BeginExpansion
{\displaystyle\sum\nolimits_{i}}
%EndExpansion
J^{(i)} \label{curre0}%
\end{equation}
which support are the set of timelike lines $\cup_{i}$ $\sigma^{(i)}$. If we
introduce a global coordinate chart for $M$ with coordinates $\{\mathrm{x}%
^{\mu}\}$ in the Einstein-Lorentz-Poincar\'{e} gauge\footnote{Given a
Minkowski spacetime structure, global coordinates \{$\mathrm{x}^{\mu}$\} for
$M\simeq\mathbb{R}^{4}$ are said to be in the Einstein-Lorentz-Poincar\'{e}
gauge (\textit{ELPG}) if and only if the following conditions hold: $%
%TCIMACRO{\TeXButton{g}{\slg}}%
%BeginExpansion
\slg
%EndExpansion
=\eta_{\mu\nu}d\mathrm{x}^{\mu}\otimes d\mathrm{x}^{\nu}$, $D_{\frac{\partial
}{\partial\mathrm{x}^{\mu}}}\frac{\partial}{\partial\mathrm{x}^{\nu}}=0$. Of
course, as well known, there exists an infinity of coordinate functions
related by Poincar\'{e} transformations satisfying these conditions. We shall
write in what follows $\gamma^{\mu}:=d\mathrm{x}^{\mu}$ and $\gamma_{\mu}%
=\eta_{\mu\nu}\gamma^{\nu}.$} then we can write
\begin{equation}
J=%
%TCIMACRO{\dsum \nolimits_{i}}%
%BeginExpansion
{\displaystyle\sum\nolimits_{i}}
%EndExpansion
J_{\mu}^{(i)}\gamma^{\mu}, \label{curre1}%
\end{equation}%
\begin{equation}
J_{\mu}^{(i)}=\eta_{\mu\nu}q^{(i)}%
%TCIMACRO{\dint }%
%BeginExpansion
{\displaystyle\int}
%EndExpansion
\delta^{(4)}(\mathrm{x}^{\beta}-\mathrm{x}^{\beta}\circ\sigma^{(i)}%
(s_{(i)}))\frac{d\mathrm{x}^{\nu}\circ\sigma^{(i)}(s_{(i)})}{ds_{(i)}}%
ds_{(i)}, \label{curre2}%
\end{equation}
with $s_{(i)}$ being the proper time along $\sigma^{(i)}$. Before going on we
must say that if the density of particles is very large we may eventually
approximate $J$ by a continuous section of $%
%TCIMACRO{\dbigwedge \nolimits^{1}}%
%BeginExpansion
{\displaystyle\bigwedge\nolimits^{1}}
%EndExpansion
T^{\ast}M$ or at least by a de Rham current \cite{derham}. It is an empirical
fact that $F$ is closed, i.e., $dF=0$ and moreover\footnote{The symbol
$\underset{%
%TCIMACRO{\TeXButton{g}{\sslg}}%
%BeginExpansion
\sslg
%EndExpansion
}{\mathbf{\star}}$ means the \textit{pair }Hodge dual operator, and its
definition is given below.}, $\mathbf{J=}\underset{%
%TCIMACRO{\TeXButton{g}{\sslg}}%
%BeginExpansion
\sslg
%EndExpansion
}{\mathbf{\star}}J\in\sec%
%TCIMACRO{\dbigwedge \nolimits^{3}}%
%BeginExpansion
{\displaystyle\bigwedge\nolimits^{3}}
%EndExpansion
T^{\ast}M$ is exact, i.e., $\mathbf{J=-}dG$ for $G\in\sec%
%TCIMACRO{\dbigwedge \nolimits^{3}}%
%BeginExpansion
{\displaystyle\bigwedge\nolimits^{3}}
%EndExpansion
T^{\ast}M$, called the excitation field. Those empirical observations are
written as
\begin{equation}
dF=0,\text{ }dG=-\mathbf{J,} \label{1}%
\end{equation}
and known as Maxwell equations.

\begin{remark}
Before we proceed it must be said that if we forget the fact that the carriers
of charges are particles and simply suppose that experimentally all we have is
a $\mathbf{J\in}\sec%
%TCIMACRO{\dbigwedge \nolimits^{3}}%
%BeginExpansion
{\displaystyle\bigwedge\nolimits^{3}}
%EndExpansion
T^{\ast}M$ that is conserved \emph{(i}.e., $d\mathbf{J=0}$\emph{)}, then
supposing that the manifold where $\mathbf{J}$ lives is star-shaped, we have
$dG=-\mathbf{J}$. Using moreover the fact that $dF=0$ \emph{(}meaning that
magnetic monopoles do not exist\emph{)} it is a mathematical fact that the
system of differential equations given by \emph{Eq.(\ref{1}) }does need for
its writing only the structure of the bare\emph{ }manifold structure $M$,
i.e., \ it does not need the additional objects $(%
%TCIMACRO{\TeXButton{slg}{\slg}}%
%BeginExpansion
\slg
%EndExpansion
,D,\tau_{{%
%TCIMACRO{\TeXButton{slg}{\sslg}}%
%BeginExpansion
\sslg
%EndExpansion
}},\uparrow)$ entering the structure of Minkowski spacetime\footnote{This has
been originally observed by Cartan in \cite{cartan0}.}. However,
electromagnetism is \emph{not }only Maxwell equations, we must yet to specify
the way that two currents $\mathbf{J}^{(1)},\mathbf{J}^{(2)}\sec%
%TCIMACRO{\dbigwedge \nolimits^{3}}%
%BeginExpansion
{\displaystyle\bigwedge\nolimits^{3}}
%EndExpansion
T^{\ast}M$ interact. And to do this we shall need to use the additional
structure, as we shall see.
\end{remark}

To proceed with our presentation of electromagnetism we must recall that as it
is well known the metric tensor can be used to give a Clifford bundle
structure to $%
%TCIMACRO{\dbigwedge }%
%BeginExpansion
{\displaystyle\bigwedge}
%EndExpansion
T^{\ast}M=%
%TCIMACRO{\dbigoplus \limits_{p=0}^{4}}%
%BeginExpansion
{\displaystyle\bigoplus\limits_{p=0}^{4}}
%EndExpansion%
%TCIMACRO{\dbigwedge \nolimits^{p}}%
%BeginExpansion
{\displaystyle\bigwedge\nolimits^{p}}
%EndExpansion
T^{\ast}M$, which will be called (for reasons to be explained below) the
\textit{pair} bundle of differential forms. The Clifford bundle of
nonhomogeneous differential forms is denoted by\footnote{Details on the
construction of $\mathcal{C\ell(}M,g)$ may be found, e.g., in
\cite{rodoliv2007}.} $\mathcal{C\ell(}M,g)$, where $g\in\sec T_{2}^{0}M$
denotes the metric of the cotangent bundle, such that for any arbitrary basis
$\{e_{\mu}\}$ of $TU\subseteq TM$ and dual basis $\{%
%TCIMACRO{\TeXButton{teta}{\mbox{\boldmath{$\theta$}}}}%
%BeginExpansion
\mbox{\boldmath{$\theta$}}%
%EndExpansion
^{\mu}\}$ of $T^{\ast}U\subseteq T^{\ast}M$ ($U$ are open sets in $M$),
$\theta^{\mu}\in\sec%
%TCIMACRO{\dbigwedge \nolimits^{1}}%
%BeginExpansion
{\displaystyle\bigwedge\nolimits^{1}}
%EndExpansion
T^{\ast}U\subseteq\sec%
%TCIMACRO{\dbigwedge \nolimits^{1}}%
%BeginExpansion
{\displaystyle\bigwedge\nolimits^{1}}
%EndExpansion
T^{\ast}M\hookrightarrow\sec\mathcal{C\ell(}M,g)$, we have $%
%TCIMACRO{\TeXButton{slg}{\slg}}%
%BeginExpansion
\slg
%EndExpansion
=g_{\mu\nu}\theta^{\mu}\otimes\theta^{\nu}$, $g=g^{\mu\nu}e_{\mu}\otimes
e_{\nu}$ and $g_{\mu\nu}g^{\nu\lambda}=\delta_{\mu}^{\lambda}$.

\begin{remark}
We recall that any section of $%
%TCIMACRO{\dbigwedge \nolimits^{r}}%
%BeginExpansion
{\displaystyle\bigwedge\nolimits^{r}}
%EndExpansion
T^{\ast}M$ is said to be a \ $r$-graded form field \emph{(}or $r$-form for
short\emph{)}. Sometimes it is said to be of \emph{even} or \emph{odd} grade,
depending on whether $r$ is even or odd. This classification is not to be
confused to the concept of de Rham \emph{pair} and \emph{impair} forms, to be
introduced below.
\end{remark}

\subsection{Energy-Momentum 1-Form and Energy-Momentum Tensor for the System
of Charged Particles}

For use in Section 6 we define now the energy-momentum 1-form for a charged
particle $(m^{(i)},q^{(i)},\sigma^{(i)})$ as the $1$-form field $p^{(i)}$ over
$\sigma^{(i)}$ given by
\begin{equation}
p^{(i)}=m^{(i)}v^{(i)}, \label{emc}%
\end{equation}
and it is obvious that $p^{(i)}\cdot p^{(i)}=(m^{(i)})^{2}$. In an inertial
frame\footnote{An inertial frame is defined as a time like vector field
$\mathbf{I}$ such that $%
%TCIMACRO{\TeXButton{g}{\slg}}%
%BeginExpansion
\slg
%EndExpansion
(\mathbf{I,I})=1$ and $D\mathbf{I=0}$. More details if need may be found,
e.g., at \cite{rodoliv2007}.} $\mathbf{I}=\frac{\partial}{\partial
\mathrm{x}^{0}}$ associated to the coordinates $\{\mathrm{x}^{\mu}\}$ for $M$
in \textit{ELPG} at time \textrm{x}$^{0}=t$, the particles will occupy
different spacetime points $(t,\mathrm{x}_{(i)}^{1}(t),\mathrm{x}_{(i)}%
^{2}(t),\mathrm{x}_{(i)}^{3}(t))$. We can define the total momentum of the
particles at time $t$ only if it is \textit{licit} to sum distinct 1-forms at
different tangent spaces of $M$. This, of course requires an \textit{absolute
parallelism} and here is then a place where the flat connection $D$ that was
introduced in the structure of Minkowski spacetime becomes necessary. It
permits us to write the total momentum of the particles at time $t$ as%
\begin{equation}
P(t)=%
%TCIMACRO{\dsum \nolimits_{i}}%
%BeginExpansion
{\displaystyle\sum\nolimits_{i}}
%EndExpansion
p^{(i)}(t), \label{tm}%
\end{equation}
a necessary concept needed in order to be possible to talk about
energy-momentum conservation for the system of particles and the
electromagnetic field (see Section 6). Besides the momentum 1-form of the
particles we shall need also to introduce the energy-momentum $1$-forms
$\mathbf{T}^{\alpha}\in\sec%
%TCIMACRO{\dbigwedge \nolimits^{1}}%
%BeginExpansion
{\displaystyle\bigwedge\nolimits^{1}}
%EndExpansion
T^{\ast}M$ \ for the system of charged particles. We have:%

\begin{align}
\mathbf{T}^{\alpha}  &  =\mathbf{T}^{\alpha\beta}\gamma_{\beta},\nonumber\\
\mathbf{T}^{\alpha\beta}  &  =%
%TCIMACRO{\dsum \nolimits_{i}}%
%BeginExpansion
{\displaystyle\sum\nolimits_{i}}
%EndExpansion
\eta^{\alpha\mu}%
%TCIMACRO{\dint }%
%BeginExpansion
{\displaystyle\int}
%EndExpansion
p_{\mu}^{(i)}(s)\frac{d}{ds}\mathrm{x}^{\beta}\circ\sigma^{(i)}(s_{(i)}%
)\delta^{4}(\mathrm{x}^{\kappa}-\mathrm{x}^{\kappa}\circ\sigma^{(i)}%
(s_{(i)}))ds_{(i)}. \label{emtm}%
\end{align}

\section{The Pair Metric\ Volume Element $\tau_{{%
%TCIMACRO{\TeXButton{slg}{\sslg}}%
%BeginExpansion
\sslg
%EndExpansion
}}$}

First introduce an arbitrary $%
%TCIMACRO{\TeXButton{slg}{\slg}}%
%BeginExpansion
\slg
%EndExpansion
$-orthonormal basis $\{\mathbf{e}_{\alpha}\}$ for $TM$ and corresponding dual
basis $\{%
%TCIMACRO{\TeXButton{teta}{\mbox{\boldmath{$\theta$}}}}%
%BeginExpansion
\mbox{\boldmath{$\theta$}}%
%EndExpansion
^{\alpha}\}$ for $T^{\ast}M$. Then, $%
%TCIMACRO{\TeXButton{slg}{\slg}}%
%BeginExpansion
\slg
%EndExpansion
(\mathbf{e}_{\alpha},\mathbf{e}_{\beta})=\eta_{\alpha\beta}$ and $g(%
%TCIMACRO{\TeXButton{teta}{\mbox{\boldmath{$\theta$}}}}%
%BeginExpansion
\mbox{\boldmath{$\theta$}}%
%EndExpansion
^{\alpha},%
%TCIMACRO{\TeXButton{teta}{\mbox{\boldmath{$\theta$}}}}%
%BeginExpansion
\mbox{\boldmath{$\theta$}}%
%EndExpansion
^{\beta})=%
%TCIMACRO{\TeXButton{teta}{\mbox{\boldmath{$\theta$}}}}%
%BeginExpansion
\mbox{\boldmath{$\theta$}}%
%EndExpansion
^{\alpha}\cdot%
%TCIMACRO{\TeXButton{teta}{\mbox{\boldmath{$\theta$}}}}%
%BeginExpansion
\mbox{\boldmath{$\theta$}}%
%EndExpansion
^{\beta}=\eta^{\alpha\beta}$ and $%
%TCIMACRO{\TeXButton{teta}{\mbox{\boldmath{$\theta$}}}}%
%BeginExpansion
\mbox{\boldmath{$\theta$}}%
%EndExpansion
^{\alpha}(\mathbf{e}_{\beta})=\delta_{\beta}^{\alpha}$, where the matrix with
entries $\eta_{\alpha\beta}$ and the matrix with entries $\eta^{\mu\nu}$ are
equal to the diagonal matrix $\mathrm{diag}(1,-1,-1,-1)$. We define a
\textit{pair} metric volume\footnote{The impair volume elements is defined in
Section 4.1.} $\tau_{{%
%TCIMACRO{\TeXButton{slg}{\sslg}}%
%BeginExpansion
\sslg
%EndExpansion
}}$ $\in\sec%
%TCIMACRO{\dbigwedge \nolimits^{4}}%
%BeginExpansion
{\displaystyle\bigwedge\nolimits^{4}}
%EndExpansion
T^{\ast}M$ by
\begin{equation}
\tau_{{%
%TCIMACRO{\TeXButton{slg}{\sslg}}%
%BeginExpansion
\sslg
%EndExpansion
}}:=%
%TCIMACRO{\TeXButton{teta}{\mbox{\boldmath{$\theta$}}}}%
%BeginExpansion
\mbox{\boldmath{$\theta$}}%
%EndExpansion
^{0}\wedge%
%TCIMACRO{\TeXButton{teta}{\mbox{\boldmath{$\theta$}}}}%
%BeginExpansion
\mbox{\boldmath{$\theta$}}%
%EndExpansion
^{1}\wedge%
%TCIMACRO{\TeXButton{teta}{\mbox{\boldmath{$\theta$}}}}%
%BeginExpansion
\mbox{\boldmath{$\theta$}}%
%EndExpansion
^{2}\wedge%
%TCIMACRO{\TeXButton{teta}{\mbox{\boldmath{$\theta$}}}}%
%BeginExpansion
\mbox{\boldmath{$\theta$}}%
%EndExpansion
^{3} \label{1'}%
\end{equation}

\textbf{Remark 1 }An \textit{orientation} for $M$ as we already said above is
a free choice of an arbitrary volume element.\footnote{Of course, an arbitrary
manifold $M$, even if orientable, when equipped with an arbitrary Lorentzian
metric field \texttt{g} does not in general admit a global \texttt{g}%
-orthonormal cotetrad field, so, in this case the introduction of
$\tau_{\mathtt{g}}$ is a little more complicated \cite{choquet,darling}.
However, all manifolds $M$ part of a Lorentzian spacetime structure that
admits spinor fields have a global \texttt{g}-orthonormal cotetrad field. This
is a result from a famous theorem due to Geroch \cite{geroch}.} Before
proceeding let us introduce arbitrary coordinates $\{x^{\mu}\}$ for $U\subset
M$ and $\{x^{{\prime}\mu}\}$ for $U^{\prime}\subset M$, $U\cap U^{\prime}%
\neq\varnothing$ such that
\begin{equation}%
%TCIMACRO{\TeXButton{teta}{\mbox{\boldmath{$\theta$}}}}%
%BeginExpansion
\mbox{\boldmath{$\theta$}}%
%EndExpansion
^{\alpha}=h_{\mu}^{\alpha}dx^{\mu}\text{, }%
%TCIMACRO{\TeXButton{teta}{\mbox{\boldmath{$\theta$}}}}%
%BeginExpansion
\mbox{\boldmath{$\theta$}}%
%EndExpansion
^{\alpha}=h_{\mu}^{\prime\alpha}dx^{\prime\mu}. \label{6bis}%
\end{equation}
Let $\mathbf{h}$ and $\mathbf{h}^{\prime}$ the matrices with entries $h_{\mu
}^{\alpha}$ and $h_{\mu}^{\prime\alpha}$. Then, e.g., \
\begin{equation}
\det(g_{\alpha\beta})=(\det\mathbf{h)}^{2}\det(\eta_{\alpha\beta}),
\label{6bb}%
\end{equation}
and%
\begin{equation}
\sqrt{\left\vert \det(g_{\alpha\beta})\right\vert }=\left\vert \det
\mathbf{h}\right\vert \sqrt{\left\vert \det(\eta_{\alpha\beta})\right\vert
}=\left\vert \det\mathbf{h}\right\vert \label{6bbb}%
\end{equation}

The expression of $\tau_{{%
%TCIMACRO{\TeXButton{slg}{\sslg}}%
%BeginExpansion
\sslg
%EndExpansion
}}$ in the bases $\{dx^{\mu}\}$ and $\{dx^{\prime\mu}\}$ are respectively
\begin{align}
\tau_{{%
%TCIMACRO{\TeXButton{slg}{\sslg}}%
%BeginExpansion
\sslg
%EndExpansion
}}  &  =\frac{1}{4!}\tau_{i_{0}\ldots i_{3}}dx^{i_{0}}\wedge\cdots\wedge
dx^{i_{3}}=\tau_{0123}dx^{0}\wedge\cdots\wedge dx^{3}\nonumber\\
&  =\frac{\det\mathbf{h}}{\left\vert \det\mathbf{h}\right\vert }%
\sqrt{\left\vert \det(g_{\alpha\beta})\right\vert }dx^{0}\wedge\cdots\wedge
dx^{3}, \label{6.4}%
\end{align}
and
\begin{equation}
\tau_{{%
%TCIMACRO{\TeXButton{slg}{\sslg}}%
%BeginExpansion
\sslg
%EndExpansion
}}=\frac{1}{4!}\tau_{j_{0}\ldots j_{3}}^{\prime}dx^{\prime j_{0}}\wedge
\cdots\wedge dx^{\prime j_{3}}=\tau_{0123}^{\prime}dx^{\prime0}\wedge
\cdots\wedge dx^{\prime3}. \label{6}%
\end{equation}

Now writing $\Lambda_{j_{p}}^{i_{p}}=\frac{\partial x^{i_{p}}}{\partial
x^{\prime j_{p}}}$, $\det\Lambda=\det(\frac{\partial x^{i}}{\partial x^{\prime
j}})$\ we have (remember that $\tau_{i_{0}\ldots i_{3}}=\varepsilon
_{i_{0}\ldots i_{3}}^{0123}\tau_{0123}=\varepsilon_{i_{0}\ldots i_{3}}%
^{0123}\frac{\det\mathbf{h}}{\left\vert \det\mathbf{h}\right\vert }%
\sqrt{\left\vert \det(g_{ij})\right\vert }$)
\begin{equation}
\tau_{j_{0}\ldots j_{3}}^{\prime}=\Lambda_{j_{0}}^{i_{0}}\ldots\Lambda_{j_{3}%
}^{i_{3}}\tau_{i_{0}\ldots i_{3}}. \label{7}%
\end{equation}
Also, since $\sqrt{\left\vert \det(g_{ij}^{\prime})\right\vert }=\left\vert
\det\Lambda\right\vert $ $\sqrt{\left\vert \det(g_{ij})\right\vert }$, we end
with
\begin{align}
\tau_{0123}^{\prime}  &  =\det\Lambda\tau_{0123}:=\Delta^{-1}\tau
_{0123}\label{8a}\\
&  =\frac{\det\mathbf{h}}{\left\vert \det\mathbf{h}\right\vert }\frac
{\det\Lambda}{\left\vert \det\Lambda\right\vert }\sqrt{\left\vert \det
(g_{ij}^{\prime})\right\vert }. \label{9a}%
\end{align}

\begin{remark}
We want to emphasize here that with the choice $\frac{\det\mathbf{h}%
}{\left\vert \det\mathbf{h}\right\vert }=+1$, the coordinate expression for
$\tau_{{%
%TCIMACRO{\TeXButton{slg}{\sslg}}%
%BeginExpansion
\sslg
%EndExpansion
}}$ in the basis $\{dx^{\mu}\}$ becomes the one appearing in almost all
textbooks, i.e., $\sqrt{\left\vert \det(g_{ij})\right\vert }dx^{0}\wedge
\cdots\wedge dx^{3}$. However the coordinate expression for $\tau_{{%
%TCIMACRO{\TeXButton{slg}{\sslg}}%
%BeginExpansion
\sslg
%EndExpansion
}}$ in the basis $\{dx^{\prime\mu}\}$ is $\sqrt{\left\vert \det(g_{ij}%
^{\prime})\right\vert }dx^{\prime0}\wedge\cdots\wedge dx^{\prime3}$ only if
$\frac{\det\Lambda}{\left\vert \det\Lambda\right\vert }=+1$. The omission of
the factor $\frac{\det\Lambda}{\left\vert \det\Lambda\right\vert }$ in the
textbook presentation of $\tau_{{%
%TCIMACRO{\TeXButton{slg}{\sslg}}%
%BeginExpansion
\sslg
%EndExpansion
}}$ is the source of a big confusion and eventually responsible for a
statement saying that the volume element must be an impair $4$-form. An
\textit{impair} volume element is an object different from $\tau_{{%
%TCIMACRO{\TeXButton{slg}{\sslg}}%
%BeginExpansion
\sslg
%EndExpansion
}}$ and will be introduced in \emph{Section 4.1}.

Comparing \emph{Eq.(\ref{8a})} to \emph{Eq.(8.1)} of Schouten's book
\emph{\cite{schouten}} we see the reason why a quantity that \textquotedblleft
transforms" like in \emph{Eq.(\ref{8a})} is called a \textit{scalar-}$\Delta
$\textit{-density of weight }$1$. Despite being fan of Schouten's book, the
authors think that such a nomenclature may induce confusion, unless expressed
in a coordinate free way as, e.g., done in \emph{\cite{abraham,bott,marmo}}.
\end{remark}

\subsection{The Pair Hodge Star Operator}

A pair metric volume element $\tau_{{%
%TCIMACRO{\TeXButton{slg}{\sslg}}%
%BeginExpansion
\sslg
%EndExpansion
}}$ permits us to define an isomorphism between $%
%TCIMACRO{\dbigwedge \nolimits^{p}}%
%BeginExpansion
{\displaystyle\bigwedge\nolimits^{p}}
%EndExpansion
T^{\ast}M\hookrightarrow\mathcal{C\ell(}M,g)$ and $%
%TCIMACRO{\dbigwedge \nolimits^{4-p}}%
%BeginExpansion
{\displaystyle\bigwedge\nolimits^{4-p}}
%EndExpansion
T^{\ast}M\hookrightarrow\mathcal{C\ell(}M,g)$, given by
\begin{align}
\underset{\tau_{{%
%TCIMACRO{\TeXButton{slg}{\sslg}}%
%BeginExpansion
\sslg
%EndExpansion
}}}{\star}:%
%TCIMACRO{\dbigwedge \nolimits^{p}}%
%BeginExpansion
{\displaystyle\bigwedge\nolimits^{p}}
%EndExpansion
T^{\ast}M  &  \rightarrow%
%TCIMACRO{\dbigwedge \nolimits^{4-p}}%
%BeginExpansion
{\displaystyle\bigwedge\nolimits^{4-p}}
%EndExpansion
T^{\ast}M\nonumber\\
A_{p}  &  \mapsto\underset{\tau_{{%
%TCIMACRO{\TeXButton{slg}{\sslg}}%
%BeginExpansion
\sslg
%EndExpansion
}}}{\star}A_{p}:=\tilde{A}_{p}\tau_{{%
%TCIMACRO{\TeXButton{slg}{\sslg}}%
%BeginExpansion
\sslg
%EndExpansion
}} \label{2}%
\end{align}

In Eq.(\ref{2}) $\tilde{A}_{p}\tau_{{%
%TCIMACRO{\TeXButton{slg}{\sslg}}%
%BeginExpansion
\sslg
%EndExpansion
}}$ means the Clifford product between the Clifford fields $\tilde{A}_{p}$ and
$\tau_{%
%TCIMACRO{\TeXButton{slg}{\sslg}}%
%BeginExpansion
\sslg
%EndExpansion
}$, and $\tilde{A}_{p}$ is the reverse\footnote{See, e.g., \cite{rodoliv2007}
for details.} of $A_{p}$. Let $\{\mathrm{x}^{\mu}\}$ be global coordinates in
the \textit{ELPG} and $\{\gamma^{\mu}=d\mathrm{x}^{\mu}\}$ an orthonormal
cobasis, i.e., $g(\gamma^{\mu},\gamma^{\nu}):=\gamma^{\mu}\cdot\gamma^{\nu
}=\eta^{\mu\nu}$.

In this case we \textit{can} write $\tau_{{%
%TCIMACRO{\TeXButton{slg}{\sslg}}%
%BeginExpansion
\sslg
%EndExpansion
}}=\gamma^{5}=\gamma^{0}\wedge\gamma^{1}\wedge\gamma^{2}\wedge\gamma
^{3}=\gamma^{0}\gamma^{1}\gamma^{2}\gamma^{3}$ and the calculation of the
action of the Hodge dual operator on a $p$-form becomes an elementary
algebraic operation\footnote{Of course, our statement is true only for someone
that knows a little bit of Clifford algebra, as it is supposed to be the case
of a reader of the present article.}. We also suppose that $\tau_{{{}%
%TCIMACRO{\TeXButton{slg}{\sslg}}%
%BeginExpansion
\sslg
%EndExpansion
}}=\gamma^{5}$ defines a \textit{positive orientation }(also called
\textit{right handed orientation}), and it is trivial to verify
that\footnote{The Clifford product in this paper is represented by
juxtaposition of symbols following the convention of \cite{rodoliv2007}.}%
\begin{equation}
\tau_{{%
%TCIMACRO{\TeXButton{slg}{\sslg}}%
%BeginExpansion
\sslg
%EndExpansion
}}\tau_{{%
%TCIMACRO{\TeXButton{slg}{\sslg}}%
%BeginExpansion
\sslg
%EndExpansion
}}=\tau_{{%
%TCIMACRO{\TeXButton{slg}{\sslg}}%
%BeginExpansion
\sslg
%EndExpansion
}}\cdot\tau_{{%
%TCIMACRO{\TeXButton{slg}{\sslg}}%
%BeginExpansion
\sslg
%EndExpansion
}}=(\gamma^{5})^{2}=-1 \label{3}%
\end{equation}
Before we proceed, recall that we can show trivially that the definition given
by Eq.(\ref{2}) is equivalent to the standard one, i.e., for any $A_{p,}%
B_{p}\in\sec%
%TCIMACRO{\dbigwedge \nolimits^{p}}%
%BeginExpansion
{\displaystyle\bigwedge\nolimits^{p}}
%EndExpansion
T^{\ast}M$, it follows that
\begin{equation}
B_{p}\wedge\underset{\tau_{{%
%TCIMACRO{\TeXButton{slg}{\sslg}}%
%BeginExpansion
\sslg
%EndExpansion
}}}{\star}A_{p}=(B_{p}\cdot A_{p})\tau_{{%
%TCIMACRO{\TeXButton{slg}{\sslg}}%
%BeginExpansion
\sslg
%EndExpansion
}} \label{2bis}%
\end{equation}

\begin{remark}
As defined, the object $A_{p}^{\prime}=\underset{\tau_{{%
%TCIMACRO{\TeXButton{slg}{\sslg}}%
%BeginExpansion
\sslg
%EndExpansion
}}}{\star}A_{p}$ is a legitimate $\emph{pair}$ form, although it depends, as
it is obvious from \emph{Eq.(\ref{2}), }of the chosen orientation $\tau_{{%
%TCIMACRO{\TeXButton{slg}{\sslg}}%
%BeginExpansion
\sslg
%EndExpansion
}}$. Some authors \emph{(}like e.g., \emph{\cite{hehl})} assert that the Hodge
star operator maps a pair form into an impair one \emph{(}see the definition
of impair forms below\emph{)}. What does this statement mean given our
definition that the \emph{(}pair\emph{)} Hodge star operator changes an even
grade form to an odd grade form \emph{(}and vice versa\emph{)? I}t means that
an impair Hodge operator that maps a pair form into an impair one can also be
defined and of course is a concept different from the one just introduced. The
impair Hodge operator will be presented and discussed in \emph{Section}
\emph{4.1}. To avoid any possible confusion on this issue, let us bethink that
there may two different Hodge operators associated with the same metric $%
%TCIMACRO{\TeXButton{g}{\slg}}%
%BeginExpansion
\slg
%EndExpansion
$. Indeed, given the metric $%
%TCIMACRO{\TeXButton{slg}{\slg}}%
%BeginExpansion
\slg
%EndExpansion
$ and a \emph{(}pair\emph{) }metric volume $4$-form $\tau_{{%
%TCIMACRO{\TeXButton{slg}{\sslg}}%
%BeginExpansion
\sslg
%EndExpansion
}}^{\prime}\neq$ $\tau_{{%
%TCIMACRO{\TeXButton{slg}{\sslg}}%
%BeginExpansion
\sslg
%EndExpansion
}}$ with $\tau_{{{}%
%TCIMACRO{\TeXButton{slg}{\sslg}}%
%BeginExpansion
\sslg
%EndExpansion
}}^{\prime2}=-1$ we may define another Hodge star operator
\begin{align}
\underset{\tau_{{{}%
%TCIMACRO{\TeXButton{slg}{\sslg}}%
%BeginExpansion
\sslg
%EndExpansion
}}^{\prime}}{\star}:%
%TCIMACRO{\dbigwedge \nolimits^{p}}%
%BeginExpansion
{\displaystyle\bigwedge\nolimits^{p}}
%EndExpansion
T^{\ast}M  &  \rightarrow%
%TCIMACRO{\dbigwedge \nolimits^{4-p}}%
%BeginExpansion
{\displaystyle\bigwedge\nolimits^{4-p}}
%EndExpansion
T^{\ast}M\nonumber\\
A_{p}  &  \mapsto\underset{\tau_{{{}%
%TCIMACRO{\TeXButton{slg}{\sslg}}%
%BeginExpansion
\sslg
%EndExpansion
}}^{\prime}}{\star}A_{p}:=\tilde{A}_{p}\tau_{{{}%
%TCIMACRO{\TeXButton{slg}{\sslg}}%
%BeginExpansion
\sslg
%EndExpansion
}}^{\prime} \label{4}%
\end{align}
Now, there are only two possibilities for $\tau_{{{}%
%TCIMACRO{\TeXButton{slg}{\sslg}}%
%BeginExpansion
\sslg
%EndExpansion
}}^{\prime}$. Either $\tau_{{{}%
%TCIMACRO{\TeXButton{slg}{\sslg}}%
%BeginExpansion
\sslg
%EndExpansion
}}^{\prime}=\tau_{{{}%
%TCIMACRO{\TeXButton{slg}{\sslg}}%
%BeginExpansion
\sslg
%EndExpansion
}}$ or $\tau_{{{}%
%TCIMACRO{\TeXButton{slg}{\sslg}}%
%BeginExpansion
\sslg
%EndExpansion
}}^{\prime}=-\tau_{{{}%
%TCIMACRO{\TeXButton{slg}{\sslg}}%
%BeginExpansion
\sslg
%EndExpansion
}}$. In the second case we say that $\tau_{{{}%
%TCIMACRO{\TeXButton{slg}{\sslg}}%
%BeginExpansion
\sslg
%EndExpansion
}}^{\prime}$ defines a \emph{negative} or \emph{left handed orientation}%
\textit{. }It is obvious that in this case we have\textit{ }%
\begin{equation}
\underset{\tau_{{{}%
%TCIMACRO{\TeXButton{slg}{\sslg}}%
%BeginExpansion
\sslg
%EndExpansion
}}^{\prime}}{\star}A_{p}:=-\underset{\tau_{{{}%
%TCIMACRO{\TeXButton{slg}{\sslg}}%
%BeginExpansion
\sslg
%EndExpansion
}}}{\star}A_{p}, \label{5}%
\end{equation}
but we insist: both $\underset{\tau_{{{}%
%TCIMACRO{\TeXButton{slg}{\sslg}}%
%BeginExpansion
\sslg
%EndExpansion
}}^{\prime}}{\star}A_{p}$ and $\underset{\tau_{{{}%
%TCIMACRO{\TeXButton{slg}{\sslg}}%
%BeginExpansion
\sslg
%EndExpansion
}}}{\star}A_{p}$ are legitimate pair $4$-forms.
\end{remark}

Following Feynman, we take the view that only the field $F$ is fundamental and
that the \ charge carriers moves in the vacuum (Lorentz vacuum). It is then
necessary to find the relation between $G$ and $F$ for the vacuum. It is an
empirical fact that once a spacetime orientation $\tau_{%
%TCIMACRO{\TeXButton{slg}{\sslg}}%
%BeginExpansion
\sslg
%EndExpansion
}$ is fixed (by arbitrary choice) we get a \textit{correct} description of
electromagnetic phenomena in vacuum\footnote{Of course, any formulation of
electrodynamics in a medium must take as true the equations in vacuum and the
properties of matter which are supposed to be described (due to the obvious
difficulties with the many body problem) \ by an approximate phenomenological
theory derived, e.g., from quantum mechanics. This is the point of view of
Feynman\cite{feynman} which we endorse.} by taking%
\begin{equation}
G:=\underset{\tau_{%
%TCIMACRO{\TeXButton{slg}{\sslg}}%
%BeginExpansion
\sslg
%EndExpansion
}}{\star}F \label{10}%
\end{equation}

\begin{remark}
Until now, we have only used pair forms in our formulation of
electromagnetism, but we call the reader's attention to the fact that if we
choose the opposite spacetime orientation $\tau_{%
%TCIMACRO{\TeXButton{slg}{\sslg}}%
%BeginExpansion
\sslg
%EndExpansion
}^{\prime}$ and $\underset{\tau_{%
%TCIMACRO{\TeXButton{slg}{\sslg}}%
%BeginExpansion
\sslg
%EndExpansion
}^{\prime}}{\star}$, we must put
\begin{equation}
G=-\underset{\tau_{%
%TCIMACRO{\TeXButton{slg}{\sslg}}%
%BeginExpansion
\sslg
%EndExpansion
}^{\prime}}{\star}F, \label{10bis}%
\end{equation}
if we want to preserve the non homogeneous Maxwell equation $dG=-\mathbf{J}$.
\end{remark}

\section{Impair Forms}

The definition of the Hodge dual leaves it clear that different orientations
(i.e., different \textit{pair} volume element forms differing by a sign)
produce duals --- in the Hodge's sense --- differing by a sign. This
elementary fact is sometimes confused with the concept of impair forms
introduced by de Rham \cite{derham}\emph{. }From a historical point of view it
must be recalled that de Rham pair and impair forms are only a modern
reformulation of objects already introduced by Weyl \cite{weyl} and then by
Schouten \cite{schouten}.

Let \ $\{e_{\mu}\}$ and $\{e_{\mu}^{\prime}\}\ $be arbitrary bases for
sections of $TU\subset TM$ and $TU^{\prime}\subset TM\ $\emph{(}$U^{\prime
}\cap U\neq\varnothing$\emph{)} and $\{\theta^{\mu}\}$ and $\{\theta
^{\prime\mu}\}$ be respectively bases for $\sec%
%TCIMACRO{\dbigwedge }%
%BeginExpansion
{\displaystyle\bigwedge}
%EndExpansion
T^{\ast}U\subset\sec%
%TCIMACRO{\dbigwedge }%
%BeginExpansion
{\displaystyle\bigwedge}
%EndExpansion
T^{\ast}M\hookrightarrow\sec\mathcal{C\ell(}M,g)$ and $\sec%
%TCIMACRO{\dbigwedge }%
%BeginExpansion
{\displaystyle\bigwedge}
%EndExpansion
T^{\ast}U^{\prime}\subset\sec%
%TCIMACRO{\dbigwedge }%
%BeginExpansion
{\displaystyle\bigwedge}
%EndExpansion
T^{\ast}M\hookrightarrow\sec\mathcal{C\ell(}M,g)$ which are respectively dual
to the bases $\{e_{\mu}\}$ and $\{e_{\mu}^{\prime}\}$. Let $\omega=\frac
{1}{4!}\omega_{i_{0}\ldots i_{3}}dx^{i_{0}}\wedge\cdots\wedge dx^{i_{3}}$ and
$\omega^{\prime}=\frac{1}{4!}\omega_{i_{0}\ldots i_{3}}^{\prime}dx^{\prime
i_{0}}\wedge\cdots\wedge dx^{\prime i_{3}}$ and let \ $\tau_{%
%TCIMACRO{\TeXButton{slg}{\sslg}}%
%BeginExpansion
\sslg
%EndExpansion
}$ be the orientation of the spacetime, which we recall is a \textit{free} choice.

\begin{definition}
\label{ori}We say that the ordered coframe basis $\{\theta^{\mu}\},$
$\{\theta^{\prime\mu}\}$ \emph{(}or simply $\omega,\omega^{\prime}$\emph{)}
are positive or right-handed oriented relative to $\tau_{%
%TCIMACRO{\TeXButton{slg}{\sslg}}%
%BeginExpansion
\sslg
%EndExpansion
}$ if
\begin{equation}
o(\omega):=-\omega\cdot\tau_{%
%TCIMACRO{\TeXButton{slg}{\sslg}}%
%BeginExpansion
\sslg
%EndExpansion
}>0,\text{ }o(\omega^{\prime})=-\omega^{\prime}\cdot\tau_{%
%TCIMACRO{\TeXButton{slg}{\sslg}}%
%BeginExpansion
\sslg
%EndExpansion
}>0, \label{rho}%
\end{equation}
and if%
\begin{equation}
o(\omega):=-\omega\cdot\tau_{%
%TCIMACRO{\TeXButton{slg}{\sslg}}%
%BeginExpansion
\sslg
%EndExpansion
}<0,\text{ }o(\omega^{\prime})=-\omega^{\prime}\cdot\tau_{%
%TCIMACRO{\TeXButton{slg}{\sslg}}%
%BeginExpansion
\sslg
%EndExpansion
}<0. \label{lho}%
\end{equation}
the bases are said to be negative or left-handed oriented.
\end{definition}

\begin{remark}
It is very important not to confuse the concept of orientation of a coframe
basis given by $o(\omega)$ with the spacetime orientation given by $\tau_{%
%TCIMACRO{\TeXButton{slg}{\sslg}}%
%BeginExpansion
\sslg
%EndExpansion
}$. But of course, the orientation of a coframe changes if it is referred to
another volume element with different orientation.
\end{remark}

\begin{remark}
Also, suppose that a given manifold $M$ is non orientable. In this case we
define the relative orientation of the basis $\{\theta^{\mu}\},$
$\{\theta^{\prime\mu}\}$ on $U\cap U^{\prime}$ by saying that they have the
same orientation if \ $\omega\cdot\omega^{\prime}>0$ and opposite orientation
if $\omega\cdot\omega^{\prime}<0$. In the following the symbol $o(\omega)$
will be used according to \emph{Definition \ref{ori}} if we are referring to
an orientable manifold. In the eventual case where we referred to a non
oriented manifold that even does not carry a metric field, $o(\omega^{\prime
})$ will mean the relative orientation of a given basis $\{\theta^{\prime\mu
}\}$ in $U\cap U^{\prime}$ relative to $\{\theta^{\mu}\}$, given by sign of
Jacobian determinant $\det\Lambda/\left\vert \det\Lambda\right\vert $.
\end{remark}

\begin{definition}
\label{impair}An impair $p$-form field $\overset{\bigtriangleup}{A_{p}}$ is an
equivalence class of pairs $(\overset{\bigtriangleup}{A}_{p}^{\omega}%
,o(\omega))$, where $\overset{\bigtriangleup}{A}_{p}^{\omega}\in\sec%
%TCIMACRO{\dbigwedge \nolimits^{p}}%
%BeginExpansion
{\displaystyle\bigwedge\nolimits^{p}}
%EndExpansion
T^{\ast}M$ \emph{a pair form}--- called the representative of $\overset
{\bigtriangleup}{A_{p}}$ on the basis $\{\theta^{\mu}\}$ --- is given by
\begin{equation}
\overset{\bigtriangleup}{A}_{p}^{\omega}=o(\omega)\frac{1}{p!}\overset
{\bigtriangleup}{A}_{i_{1}\ldots i_{p}}\theta^{i_{1}}\wedge\cdots\wedge
\theta^{i_{p}} \label{11}%
\end{equation}
Given the pairs $(\overset{\bigtriangleup}{A}_{p}^{\omega},o(\omega))$ and
$(\overset{\bigtriangleup}{A}_{p}^{\omega^{\prime}},o(\omega^{\prime}))$ where
$\overset{\bigtriangleup}{A}_{p}^{\omega^{\prime}}\in\sec%
%TCIMACRO{\dbigwedge \nolimits^{p}}%
%BeginExpansion
{\displaystyle\bigwedge\nolimits^{p}}
%EndExpansion
T^{\ast}M$ is given by
\begin{equation}
\overset{\bigtriangleup}{A}_{p}^{\omega^{\prime}}=o(\omega^{\prime})\frac
{1}{p!}\overset{\bigtriangleup}{A^{\prime}}_{j_{1}\ldots i_{p}}\theta%
%TCIMACRO{\U{b4}}%
%BeginExpansion
\acute{}%
%EndExpansion
^{j_{1}}\wedge\cdots\wedge\theta%
%TCIMACRO{\U{b4}}%
%BeginExpansion
\acute{}%
%EndExpansion
^{j_{p}} \label{11bis}%
\end{equation}
we say that they are equivalent if one of the two cases holds:
\begin{equation}%
\begin{array}
[c]{ccc}%
\text{\emph{(a)}} & \text{if }o(\omega)=o(\omega^{\prime}), & \overset
{\bigtriangleup}{A}_{p}^{\omega}=\text{ }\overset{\bigtriangleup}{A}%
_{p}^{\omega^{\prime}}\\
\text{\emph{(b)}} & \text{if }o(\omega)=-o(\omega^{\prime}), & \overset
{\bigtriangleup}{A}_{p}^{\omega}=-\overset{\bigtriangleup}{A}_{p}%
^{\omega^{\prime}}%
\end{array}
\label{11biss}%
\end{equation}
\ 
\end{definition}

\begin{remark}
Let $\{e_{\mu}=\frac{\partial}{\partial\mathrm{x}^{\mu}}\}$ and $\{e_{\mu
}^{\prime}=\frac{\partial}{\partial x^{^{\prime}\mu}}\}$ be coordinate bases
where $\{\mathrm{x}^{\mu}\}$ are global coordinates in the Einstein
Lorentz-Poincar\'{e} gauge for $U\subset M$ and $\{x^{\mu}\}$ coordinates for
$U^{\prime}\subset M$. Now the orientation of $\{\gamma^{\mu}=d\mathrm{x}%
^{\mu}\}$ being taken as \emph{positive}, if we \emph{simply} write \emph{(}as
did de Rham\emph{)}%

\begin{align}
\overset{\bigtriangleup}{A}_{p}^{\omega}  &  =\frac{1}{p!}\overset
{\blacktriangle}{A}_{i_{1}\ldots i_{p}}d\mathrm{x}^{i_{1}}\wedge\cdots\wedge
d\mathrm{x}^{i_{p}},\label{12}\\
\overset{\bigtriangleup}{A}_{p}^{\omega^{\prime}}  &  =\frac{1}{p!}%
\overset{\blacktriangle}{A%
%TCIMACRO{\U{b4}}%
%BeginExpansion
\acute{}%
%EndExpansion
}_{j_{1}\ldots j_{p}}dx^{\prime j_{1}}\wedge\cdots\wedge dx^{\prime j_{p}},
\end{align}
then we must have%
\begin{equation}
\overset{\blacktriangle}{A%
%TCIMACRO{\U{b4}}%
%BeginExpansion
\acute{}%
%EndExpansion
}_{j_{1}\ldots j_{p}}=\frac{\det\Lambda}{\left\vert \det\Lambda\right\vert
}\Lambda_{j_{1}}^{i_{1}}\ldots\Lambda_{j_{p}}^{i_{p}}\overset{\blacktriangle
}{A}_{i_{1}..i_{p},} \label{13}%
\end{equation}
with $\det\Lambda=\det(\frac{\partial\mathrm{x}^{i}}{\partial x^{\prime j}})$.
\emph{Eq. (\ref{13})} is the definition of an \textit{impair} form given by de
Rham\emph{ \cite{derham}} \emph{(}see also \emph{\cite{schouten})}.
\end{remark}

\begin{remark}
It is very important to note that, since according to \emph{Definition 6} an
impair $p$-form is an equivalence class of pairs $(\overset{\bigtriangleup}%
{A}_{p}^{\omega},o(\omega))$ where each $\overset{\bigtriangleup}{A}%
_{p}^{\omega}$ is a pair $p-$form and $o(\omega)$ denotes the basis
orientation, Recall that if spacetime is oriented and we define $o(\omega)$
\ by \emph{Eqs.(\ref{rho})} and \emph{(\ref{lho}), }then it depends on the
spacetime orientation $\tau_{%
%TCIMACRO{\TeXButton{slg}{\sslg}}%
%BeginExpansion
\sslg
%EndExpansion
}$, and it follows that each pair $p$-form representative of an impair
$p$-form depends also on the choice of the spacetime orientation. Indeed, if
we change the spacetime orientation from $\tau_{%
%TCIMACRO{\TeXButton{slg}{\sslg}}%
%BeginExpansion
\sslg
%EndExpansion
}$ to $\tau_{%
%TCIMACRO{\TeXButton{slg}{\sslg}}%
%BeginExpansion
\sslg
%EndExpansion
}^{\prime}=-\tau_{%
%TCIMACRO{\TeXButton{slg}{\sslg}}%
%BeginExpansion
\sslg
%EndExpansion
}$ the orientation of the coframe $\{\theta^{\mu}\}$ changes to $o^{\prime
}(\omega)=-\omega\cdot\tau_{%
%TCIMACRO{\TeXButton{slg}{\sslg}}%
%BeginExpansion
\sslg
%EndExpansion
}^{\prime}=-o(\omega)$.
\end{remark}

We denote the bundle of impair $p$-forms by $%
%TCIMACRO{\dbigwedge \nolimits_{-}^{p}}%
%BeginExpansion
{\displaystyle\bigwedge\nolimits_{-}^{p}}
%EndExpansion
T^{\ast}M$ and the exterior bundle $%
%TCIMACRO{\dbigwedge _{-}}%
%BeginExpansion
{\displaystyle\bigwedge_{-}}
%EndExpansion
T^{\ast}M=%
%TCIMACRO{\dbigoplus \limits_{p=0}^{4}}%
%BeginExpansion
{\displaystyle\bigoplus\limits_{p=0}^{4}}
%EndExpansion%
%TCIMACRO{\dbigwedge \nolimits_{-}^{p}}%
%BeginExpansion
{\displaystyle\bigwedge\nolimits_{-}^{p}}
%EndExpansion
T^{\ast}M$. Let $\overset{\bigtriangleup}{A_{p}}\in\sec%
%TCIMACRO{\dbigwedge \nolimits_{-}^{p}}%
%BeginExpansion
{\displaystyle\bigwedge\nolimits_{-}^{p}}
%EndExpansion
T^{\ast}M$ denote that the impair $p$-form field $\overset{\bigtriangleup}%
{A}_{p}$ is a section of $%
%TCIMACRO{\dbigwedge \nolimits_{-}^{p}}%
%BeginExpansion
{\displaystyle\bigwedge\nolimits_{-}^{p}}
%EndExpansion
T^{\ast}M$.

\begin{remark}
\label{line}We can easily show that $%
%TCIMACRO{\dbigwedge _{-}}%
%BeginExpansion
{\displaystyle\bigwedge_{-}}
%EndExpansion
T^{\ast}M$ \ as defined above is isomorphic to $L(M)\otimes%
%TCIMACRO{\dbigwedge }%
%BeginExpansion
{\displaystyle\bigwedge}
%EndExpansion
T^{\ast}M$ \emph{(}whose sections are line-bundle-valued multiforms\ on
$M$\emph{)} We write \emph{\cite{abraham,bott}}
\begin{equation}%
%TCIMACRO{\dbigwedge \nolimits_{-}}%
%BeginExpansion
{\displaystyle\bigwedge\nolimits_{-}}
%EndExpansion
T^{\ast}M\simeq L(M)\otimes%
%TCIMACRO{\dbigwedge }%
%BeginExpansion
{\displaystyle\bigwedge}
%EndExpansion
T^{\ast}M, \label{lb}%
\end{equation}
where $L(M)$ is the so called orientation line bundle of $M$, a vector bundle
with typical fiber $\mathbb{R}$ and where the transition functions are defined
as follows. Let $\{(U_{\alpha},\varphi_{\alpha})\}$be a coordinate covering of
$M$ with transition functions given by $t_{\alpha\beta}=\varphi_{\alpha}%
\circ\varphi_{\beta}^{-1}$. Then, the transition functions of $L(M)$ are given
by $J(t_{\alpha\beta})/\left\vert J(t_{\alpha\beta})\right\vert $, where
$J(t_{\alpha\beta})$ means the Jacobian of matrix of the partial derivatives
of $t_{\alpha\beta}$. Under the above conditions we can write \emph{(}with the
usual abuse of notation\emph{)} for a given $\overset{\triangle}{A}\in
\sec(L(M)\otimes%
%TCIMACRO{\dbigwedge }%
%BeginExpansion
{\displaystyle\bigwedge}
%EndExpansion
T^{\ast}M)$,%
\begin{equation}
\overset{\triangle}{A}=e_{(\alpha)}\otimes\overset{}{A}_{(\alpha)}=e_{(\beta
)}\otimes\overset{}{A}_{(\beta)} \label{xq}%
\end{equation}
This formula leaves it clear once again that to start any game with impair
forms we must,\ once we choose a given chart $(U_{\alpha},\varphi_{\alpha})$,
to give by \emph{convention }an orientation $e_{(\alpha)}$ for it and next we
must choose a pair form $A_{(\alpha)}=A$ or its negative, i.e., $A_{(\alpha
)}=-A$ to build $\overset{\triangle}{A}$. This choice depends of course on the
applications we have in mind.
\end{remark}

\subsection{The Impair Volume Element}

The impair $4$-form $\overset{\bigtriangleup}{\tau}_{%
%TCIMACRO{\TeXButton{sg}{\sslg}}%
%BeginExpansion
\sslg
%EndExpansion
}$ $\in\sec%
%TCIMACRO{\dbigwedge \nolimits_{-}^{4}}%
%BeginExpansion
{\displaystyle\bigwedge\nolimits_{-}^{4}}
%EndExpansion
T^{\ast}M$ whose representative\ in an arbitrary basis $\{dx^{\mu}%
\}$\ supposed positive is given by
\begin{align}
\overset{\bigtriangleup}{\tau}_{%
%TCIMACRO{\TeXButton{sg}{\sslg}}%
%BeginExpansion
\sslg
%EndExpansion
}  &  =\frac{1}{4!}\overset{\blacktriangle}{\tau}_{i_{o}i_{1}i_{2}i_{3}%
}dx^{i_{1}}\wedge\cdots\wedge dx^{i_{p}}=\overset{\blacktriangle}{\tau}%
_{0123}dx^{0}\wedge\cdots\wedge dx^{3}\nonumber\\
&  =\sqrt{\left\vert \det(g_{ij})\right\vert }dx^{0}\wedge\cdots\wedge dx^{3},
\label{13aa}%
\end{align}
is sometimes called (see, e.g., \cite{frankel}) the \textit{pseudo volume
element\footnote{Here it becomes obvious what we said in Remark \ref{line}. We
want of course, that the volume of a compact region $U\subset M$ be a positive
number. This implies that we must choose as elements in the trivialization of
$L(M)\otimes%
%TCIMACRO{\dbigwedge }%
%BeginExpansion
{\displaystyle\bigwedge}
%EndExpansion
T^{\ast}M$, $1\otimes\sqrt{\left\vert \det(g_{ij})\right\vert }dx^{0}%
\wedge\cdots\wedge dx^{3}$ or $(-1)\otimes(-\sqrt{\left\vert \det
(g_{ij})\right\vert }dx^{0}\wedge\cdots\wedge dx^{3})$ which is simply written
as $\overset{\bigtriangleup}{\tau}_{%
%TCIMACRO{\TeXButton{sg}{\sslg}}%
%BeginExpansion
\sslg
%EndExpansion
}=$ $\sqrt{\left\vert \det(g_{ij})\right\vert }dx^{0}\wedge\cdots\wedge
dx^{3}$.}}. Now, the representative of this impair form in the basis
$\{dx^{\prime\mu}\}$ is according to the definition just given
\begin{align}
\overset{\bigtriangleup}{\tau}_{%
%TCIMACRO{\TeXButton{sg}{\sslg}}%
%BeginExpansion
\sslg
%EndExpansion
}^{\prime}  &  =\frac{1}{4!}\overset{\blacktriangle}{\tau}_{i_{o}i_{1}%
i_{2}i_{3}}^{\prime}dx^{\prime i_{1}}\wedge\cdots\wedge dx^{\prime i_{p}%
}=\overset{\blacktriangle}{\tau}_{0123}^{\prime}dx^{\prime0}\wedge\cdots\wedge
dx^{\prime3}\nonumber\\
&  =\sqrt{\left\vert \det(g_{ij}^{\prime})\right\vert }dx^{\prime0}%
\wedge\cdots\wedge dx^{\prime3}, \label{13b}%
\end{align}
where we used \ that \ $\sqrt{\left\vert \det(g_{ij}^{\prime})\right\vert
}=\left\vert \det\Lambda\right\vert $ $\sqrt{\left\vert \det(g_{ij}%
)\right\vert }$ and Eq.(\ref{13}), i.e.,
\begin{equation}
\overset{\blacktriangle}{\tau^{\prime}}_{0123}=\left\vert \det\Lambda
\right\vert \overset{\blacktriangle}{\tau}_{0123}. \label{14}%
\end{equation}
Note that Eq.(\ref{14}) is different from Eq.(\ref{8a})\emph{ }which defines
the transformation rule \emph{for the components of a pair volume element.}

We recall also that given a chart $(U,\varphi)$ of the atlas of $M$, the
integral of an impair $n$-form $\overset{\bigtriangleup}{\tau}_{%
%TCIMACRO{\TeXButton{sg}{\sslg}}%
%BeginExpansion
\sslg
%EndExpansion
}$ on a compact region $R\subset U\subset M$ is according to de Rham's
definition (with $\mathbf{R = }\varphi(R)$) given by%
\begin{equation}
\int_{R}\overset{\bigtriangleup}{\tau}_{%
%TCIMACRO{\TeXButton{sg}{\sslg}}%
%BeginExpansion
\sslg
%EndExpansion
}:=\int_{\mathbf{R}}\overset{\blacktriangle}{\tau}_{0123}dx^{0}dx^{1}%
dx^{2}dx^{3}, \label{15}%
\end{equation}
and as a result of \emph{Eq.(\ref{13})} we have in the chart $(U^{\prime
},\varphi^{\prime})$, $R\subset U^{\prime}$ and with $\mathbf{R}^{\prime
}\mathbf{=}\varphi^{\prime}(R),$ $\ $%
\begin{align}
\int_{R}\overset{\bigtriangleup}{\tau}_{%
%TCIMACRO{\TeXButton{sg}{\sslg}}%
%BeginExpansion
\sslg
%EndExpansion
}^{\prime}  &  =\int_{\mathbf{R}^{\prime}}\left\vert \det\Lambda\right\vert
\overset{\blacktriangle}{\tau}_{0123}dx^{\prime0}dx^{\prime1}dx^{\prime
2}dx^{\prime3}\nonumber\\
&  =\int_{\mathbf{R}}\overset{\blacktriangle}{\tau}_{0123}dx^{0}dx^{1}%
dx^{2}dx^{3}, \label{16}%
\end{align}
which corresponds to the classical formula for variables change in a multiple
integration. Thus the integral of an impair $n$-form on a $n$-dimensional
manifold is independent of the orientation of $R$. This is \textit{not} the
case if we try to integrate a pair $n$-form. We briefly recall de Rham's
theory \cite{derham} of how to integrate pair and impair $p$-forms living on
an $n$-dimensional manifold $M$.

\begin{remark}
\label{integration}Suppose we assign the natural orientation to a `rectangle'
$U^{p}\subset\mathbb{R}^{p}$ by $\tau=d\mathbf{x}^{1}\wedge\cdots\wedge
d\mathbf{x}^{p}$ \emph{(}where $\{\mathbf{x}^{i}\}$ are Cartesian coordinates
for $\mathbb{R}^{p}$\emph{)}. It is now a classical result due to de Rham that
it is always possible to integrate a pair $p$-form $\alpha\in\sec
\bigwedge\nolimits^{p}T^{\ast}M$ over an \emph{inner} oriented $p$-chain,
i.e., a parametrized submanifold $\psi:U^{p}\rightarrow$\ $M$ endowed with an
inner orientation \emph{\cite{choquet,darling,derham,frankel}} $\bar{o}%
(U^{p})$. Indeed, if $\{u^{i}\}$ are arbitrary coordinates covering
$\psi(U^{p})$, we have by definition%
\begin{align}
\int\nolimits_{(\psi(U^{r}),\text{ }\bar{o}(U^{r}))}\alpha &  =\int
\nolimits_{(U^{r},\text{ }\bar{o}(U^{r}))}\psi^{\ast}\alpha\nonumber\\
&  :=\bar{o}(U^{r})\int\nolimits_{(U^{r},\text{ }\bar{o}(U^{r}))}\psi^{\ast
}\alpha\left(  \frac{\partial}{\partial u^{1}},\ldots,\frac{\partial}{\partial
u^{r}}\right)  du^{1}\ldots du^{r},
\end{align}
where $\bar{o}(U^{r})=\mathrm{sign}\det(\frac{\partial u^{i}}{\partial
\mathbf{x}^{j}})$. Of course, if we assign a different orientation $\bar
{o}^{\prime}(U^{r})=-\bar{o}(U^{r})$ to $U^{p}$ we have%
\begin{equation}
\int\nolimits_{(\psi(U^{r}),-\bar{o}^{\prime}(U^{r}))}\alpha=-\int
\nolimits_{(\psi(U^{r}),\text{ }\bar{o}(U^{r}))}\alpha.
\end{equation}
It is now opportune to bethink that any impair $n$-form is always integrable
over any compact $n$-dimensional manifold $M$, be it orientable or not.
However, it is \emph{not} always possible to integrate an impair $r$-form
$\overset{\bigtriangleup}{\alpha}$\ on a $n$-dimensional manifold over a
parametrized submanifold $\psi:U^{r}\rightarrow$ $M$ unless $\psi$ is an outer
orientable map, i.e., if we can associate an orientation to $M$ on $\psi
(U^{r})$. In general it may be not be possible to do that, and thus, we cannot
integrate $\overset{\bigtriangleup}{\alpha}$ over an orientable $p$%
-dimensional submanifold $S$ $\subset\psi(U^{r})$, unless $S$ is endowed with
an \emph{outer }or \emph{transverse }orientation, i.e., if at any point of
$x\in V$, $T_{x}M=$ $T_{x}S\oplus N$ \emph{(}with any $n\in N$ being
transverse to $S$, i.e., $n\notin T_{x}S$\emph{)} then each transversal $N$
can be oriented continuously as a function of $x\in V$. Let $(x^{1}%
,\ldots,x^{n})$ be coordinates covering $U\subset M$ such that $S\cap U$ is
defined by $x^{\iota}=f(x^{1},\ldots,x^{r})$, $\iota=r+1,\ldots,n$. Of, course
the vector fields $\frac{\partial}{\partial x^{\iota}}$, $\iota=r+1,\ldots,n$,
defined in $U$, are transverse to $S.$ Given an orientation for $S\cap U$,
there always exists a set of vector fields $\{\mathbf{e}_{1},\ldots
,\mathbf{e}_{r}\}$ in $T(S\cap U)$ that are positively oriented there, which
can be extended to all $TU$ by trivially keeping their components constant
when moving out of $V$. In this way an \emph{outer} orientation can be defined
in $U$ by saying that $\{\mathbf{e}_{1},\ldots,\mathbf{e}_{r},\frac{\partial
}{\partial x^{r+1}},\ldots,\frac{\partial}{\partial x^{n}}\}$ defines, let us
say, the positive orientation on $U$. Then, if $\overset{\bigtriangleup}{\eta
}\in\sec\bigwedge\nolimits^{r}T^{\ast}U$, the pullback form $i^{\ast}%
\overset{\bigtriangleup}{\eta}$ on $S\cap U$ \emph{(}where of course
$i:S\rightarrow M$ is the embedding of the submanifold $S$ on $M$\emph{)} is
well defined, and can be integrated \emph{\cite{frankel}}. Finally, recall
that the orientations defined by $o(\omega)$ and $\bar{o}(U^{r})$ are
obviously related, and we do not need further explanation.
\end{remark}

\begin{remark}
It is essential to recall that $%
%TCIMACRO{\dbigwedge \nolimits_{-}}%
%BeginExpansion
{\displaystyle\bigwedge\nolimits_{-}}
%EndExpansion
T^{\ast}M$ is not \emph{closed }under the operation of exterior multiplication
and indeed to have a closed algebra \emph{(}with that product\emph{)} we need
to take into account that the exterior multiplication of forms of the same
parity is always a pair form and the exterior multiplication of forms of
different parities is always an impair form. Also, the scalar product of forms
of the same parity gives a pair $0$-form and the scalar product of forms of
different parities gives an impair $0$-form. Moreover, the differential
operator $d$ preserves the parity of forms.
\end{remark}

\begin{remark}
If we insist in using pair and impair forms for formulating the
\emph{differential }equations of motion of a given theory, e.g., in a
formulation of electromagnetism in an orientable spacetime \emph{(}something
that \ at this point the reader must be convinced that it is not necessary at
all, as it clear from the presentation given above\emph{)} we need to
introduce an impair Hodge star operator.
\end{remark}

\subsection{The Impair Hodge Star Operator}

\begin{definition}
Let $\overset{\bigtriangleup}{\tau}_{{{}%
%TCIMACRO{\TeXButton{slg}{\sslg}}%
%BeginExpansion
\sslg
%EndExpansion
}}$ $\in\sec%
%TCIMACRO{\dbigwedge \nolimits_{-}^{4}}%
%BeginExpansion
{\displaystyle\bigwedge\nolimits_{-}^{4}}
%EndExpansion
T^{\ast}M$ be an impair volume form. The impair Hodge star operator is the map%

\begin{align}
\underset{\overset{\bigtriangleup}{\tau}_{%
%TCIMACRO{\TeXButton{slg}{\sslg}}%
%BeginExpansion
\sslg
%EndExpansion
}}{\star}  &  :%
%TCIMACRO{\dbigwedge \nolimits^{p}}%
%BeginExpansion
{\displaystyle\bigwedge\nolimits^{p}}
%EndExpansion
T^{\ast}M\rightarrow%
%TCIMACRO{\dbigwedge \nolimits_{-}^{4-p}}%
%BeginExpansion
{\displaystyle\bigwedge\nolimits_{-}^{4-p}}
%EndExpansion
T^{\ast}M,\label{17}\\
\underset{\overset{\bigtriangleup}{\tau}_{%
%TCIMACRO{\TeXButton{slg}{\sslg}}%
%BeginExpansion
\sslg
%EndExpansion
}}{\star}  &  :%
%TCIMACRO{\dbigwedge \nolimits_{-}^{p}}%
%BeginExpansion
{\displaystyle\bigwedge\nolimits_{-}^{p}}
%EndExpansion
T^{\ast}M\rightarrow%
%TCIMACRO{\dbigwedge \nolimits^{4-p}}%
%BeginExpansion
{\displaystyle\bigwedge\nolimits^{4-p}}
%EndExpansion
T^{\ast}M
\end{align}
such that for any $A_{p}\in\sec%
%TCIMACRO{\dbigwedge \nolimits^{p}}%
%BeginExpansion
{\displaystyle\bigwedge\nolimits^{p}}
%EndExpansion
T^{\ast}M$ and $\overset{\bigtriangleup}{B_{p}}\in\sec%
%TCIMACRO{\dbigwedge \nolimits_{-}^{p}}%
%BeginExpansion
{\displaystyle\bigwedge\nolimits_{-}^{p}}
%EndExpansion
T^{\ast}M$, we have
\begin{align}
\underset{\overset{\bigtriangleup}{\tau}_{{{}%
%TCIMACRO{\TeXButton{slg}{\sslg}}%
%BeginExpansion
\sslg
%EndExpansion
}}}{\star}\overset{}{A_{p}}  &  :=\overset{}{\tilde{A}_{p}}\overset
{\bigtriangleup}{\tau}_{{{}%
%TCIMACRO{\TeXButton{slg}{\sslg}}%
%BeginExpansion
\sslg
%EndExpansion
}}^{\omega},\nonumber\\
\underset{\overset{\bigtriangleup}{\tau}_{{{}%
%TCIMACRO{\TeXButton{slg}{\sslg}}%
%BeginExpansion
\sslg
%EndExpansion
}}}{\star}\overset{\bigtriangleup}{B_{p}}  &  :=\overset{\bigtriangleup
}{\tilde{B}_{p}}^{\omega}\overset{\bigtriangleup}{\tau}_{{%
%TCIMACRO{\TeXButton{slg}{\sslg}}%
%BeginExpansion
\sslg
%EndExpansion
}}^{\omega}. \label{18}%
\end{align}

\end{definition}

Note that in Eq.(\ref{18}) the Clifford product for the representatives of the
impair forms is well defined (since according to Definition \ref{impair}
\ each representative is an \textit{even} form). Given the existence of impair
and pair forms, many authors, e.g., \emph{\cite{frankel,post, post1,post2,
kiehn}} advocate that even in an orientable manifold the formulation of
electromagnetism must necessarily use besides the pair field strength
$F\in\sec%
%TCIMACRO{\dbigwedge \nolimits^{2}}%
%BeginExpansion
{\displaystyle\bigwedge\nolimits^{2}}
%EndExpansion
T^{\ast}M$ an impair exact $3$-form $\overset{\bigtriangleup}{\mathbf{J}}%
\in\sec%
%TCIMACRO{\dbigwedge \nolimits_{-}^{3}}%
%BeginExpansion
{\displaystyle\bigwedge\nolimits_{-}^{3}}
%EndExpansion
T^{\ast}M$, which then defines the excitation field as an impair $2$-form
$\overset{\bigtriangleup}{G}\in\sec%
%TCIMACRO{\dbigwedge \nolimits_{-}^{3}}%
%BeginExpansion
{\displaystyle\bigwedge\nolimits_{-}^{3}}
%EndExpansion
T^{\ast}M$. We have thus for the vacuum situation
\begin{align}
dF  &  =0,\text{ }d\overset{\bigtriangleup}{G}\text{ }=-\overset
{\bigtriangleup}{\mathbf{J}},\nonumber\\
\overset{\bigtriangleup}{G}\text{ }  &  =\underset{\overset{\bigtriangleup
}{\tau_{%
%TCIMACRO{\TeXButton{slg}{\sslg}}%
%BeginExpansion
\sslg
%EndExpansion
}}}{\star}F. \label{19}%
\end{align}

\begin{remark}
\label{re16}Authors \emph{\cite{frankel,post, post1,post2, kiehn}}
\emph{(}since the classical presentation of \emph{\cite{scho}) }offers as the
main argument for the necessity of using an impair $\overset{\bigtriangleup
}{\mathbf{J}}\in\sec%
%TCIMACRO{\dbigwedge \nolimits_{-}^{3}}%
%BeginExpansion
{\displaystyle\bigwedge\nolimits_{-}^{3}}
%EndExpansion
T^{\ast}M$ in electromagnetism the following statement \emph{(}which takes
into account the \emph{Remark} \emph{\ref{integration})} : \textquotedblleft
the value of the charge%
\begin{equation}
Q=\int\nolimits_{S}\overset{\bigtriangleup}{\mathbf{J}} \label{19'}%
\end{equation}
\ contained in a compact spacelike hypersurface $S\subset M$ must be
independent of the orientation of $S$\textquotedblright\emph{\footnote{Once we
use impair forms, following \cite{frankel} we may say that charge is a scalar.
However, take care, in, e.g., \cite{kiehn} charge is said to be a
pseudo-scalar. This apparent confusion comes out because in \cite{kiehn} it is
discussed the properties of charge and of other electromagnetic quantities
under a active parity operation and time reversal operators interpreted as
appropriated mappings $\mathbf{p:}M\rightarrow M$ and $\mathbf{t:}M\rightarrow
M$, where $M\simeq\mathbb{R}^{4}$ is the manifold entering in the structure of
Minkowski spacetime. In that case we can show, e.g., that if $\overset
{\bigtriangleup}{\mathbf{J}}$ is an \textit{impair} $2$-form than the pullback
form $\overset{\bigtriangleup}{\mathbf{J}}^{\prime}=$ $\mathbf{p}^{\ast
}\overset{\bigtriangleup}{\mathbf{J}}=-\overset{\bigtriangleup}{\mathbf{J}}$
and thus if $Q=\int\overset{\bigtriangleup}{\mathbf{J}}$ $\ $it follows that
$\int\overset{\bigtriangleup}{\mathbf{J}}^{\prime}=-Q$, and this according to
\cite{kiehn} justifies calling charge a pseudo-scalar. \ As we see a confusion
of tongues are also present in our subject. We are going to discuss more
details of this particular issue in another publication.} }and indeed, taking
into account that \ $\overset{\bigtriangleup}{\mathbf{J}}=\underset
{\overset{\bigtriangleup}{\tau_{%
%TCIMACRO{\TeXButton{slg}{\sslg}}%
%BeginExpansion
\sslg
%EndExpansion
}}}{\star}J$, where $J$ is given by \emph{Eq.(\ref{curre1})} we must have%
\begin{equation}
\int\nolimits_{S}\overset{\bigtriangleup}{\mathbf{J}}=%
%TCIMACRO{\dsum \nolimits_{i}}%
%BeginExpansion
{\displaystyle\sum\nolimits_{i}}
%EndExpansion
q^{(i)} \label{19''}%
\end{equation}

However, it is our view that the above argument is not a solid one. First,
there is no empirical evidence that any spacelike surface $S\subset M$ where a
real current is integrated does not possess an inner orientation that may be
make consistent with the orientation of $M$. Thus, empirical evidence asserts
that we may restrict ourself to only positively oriented charts \emph{(}see
also \emph{Section 4.3)}. But, even if we do not want to restrict ourselves to
the use of positively oriented charts we must not forget that to perform the
integral $\int\nolimits_{S}\overset{\bigtriangleup}{\mathbf{J}}$ using a chart
$(U,\varphi)$, $S\subset U$ with coordinates $\{x^{\mu}\}$ we must choose an
orientation \emph{(r}ecall \emph{Remark \ref{integration}) }for that chart and
pick a specific choice for the pair form representing $\overset{\bigtriangleup
}{\mathbf{J}}$. Suppose we make the convention that the orientation of
$dx^{0}\wedge dx^{1}\wedge dx^{2}\wedge dx^{3}$ is positive. At our disposal
there are $\mathbf{J\ }$and $-\mathbf{J}$. Which one to choose? The answer is
obvious, we choose $\mathbf{J}$, for in this case we will have $\int
\nolimits_{S}\overset{\bigtriangleup}{\mathbf{J}}=%
%TCIMACRO{\dsum \nolimits_{i}}%
%BeginExpansion
{\displaystyle\sum\nolimits_{i}}
%EndExpansion
q^{(i)}$. This means that the attribution of the charge parameters to
particles need in order to define the current depends on a \emph{convention},
the one described above.

What happens if we represent the current by a pair form $\mathbf{J}\in\sec%
%TCIMACRO{\dbigwedge \nolimits^{3}}%
%BeginExpansion
{\displaystyle\bigwedge\nolimits^{3}}
%EndExpansion
T^{\ast}M$? In this case the integral%
\begin{equation}
\int\nolimits_{S}\mathbf{J,} \label{19a}%
\end{equation}
does depend on the orientation of the chart used for its calculation. Suppose
that as in the case of the integration of the impair form we use a chart
$(U,\varphi)$, $S\subset U$ with coordinates $\{x^{\mu}\}$. Which orientation
should we give to $dx^{0}\wedge dx^{1}\wedge dx^{2}\wedge dx^{3}$? The answer
is obvious. We must choose an orientation \emph{(}the positive one\emph{)}
such that%
\begin{equation}
\int\nolimits_{S}\mathbf{J}=%
%TCIMACRO{\dsum \nolimits_{i}}%
%BeginExpansion
{\displaystyle\sum\nolimits_{i}}
%EndExpansion
q^{(i)}. \label{19A1}%
\end{equation}
So, in both cases \emph{(}use of impair or pair current forms\emph{) }to
\emph{start }\textit{the evaluation process we need} to make a convention in
order to fix the charge parameters of the charges that enter the definition of
current such that \emph{Eq.(\ref{19''}) }and \emph{Eq.(\ref{19A1}) }are true.

Now, what happens if someone using pair forms decides to calculate
$\int\nolimits_{S}\mathbf{J}$ using a chart with a different orientation than
the one\ previously used to fix the charge parameters of the particles? The
answer is that he will find that the value of the integral given by
\emph{Eq.(\ref{19a})} will be now $-%
%TCIMACRO{\dsum \nolimits_{i}}%
%BeginExpansion
{\displaystyle\sum\nolimits_{i}}
%EndExpansion
q^{(i)}$? Is this a puzzle? Of course not, the value is negative because he is
using a \emph{different }convention than the previous one.

What we should ask is that if this break of convention changes the physics of
electromagnetic phenomena? No, what happens is only that what was called a
positive charge will now be called a negative charge and what was called a
negative charge will now be called a positive one. No empirical fact will
change, only some names. This is so because this change of names does not
change any prediction of the theory concerning the motion of charged particles
that besides the coupling parameter $q$ also carries a coupling parameter $m$.
We shall explicitly demonstrate this statement in \emph{Section 6} after we
introduce the Clifford bundle formulation of electromagnetism. Here we
emphasize again: the sign of a charge can be given meaning only by making it
to interact with a charge that has been defined by convention as being
positive \emph{(}or negative\emph{)}\footnote{This point is well discussed in
\cite{tonti0,tonti} where the author uses an interesting homological approach
in the formulation of the electromagnetic laws.}.
\end{remark}

\subsection{Teaching Aliens What is Right Hand and What is Left Hand}

The previous considerations of Remark \ref{re16} are valid only if the
universe we live does not have regions composed of what we here called
antimatter. Indeed, let us recall one of Feynman's stories (at page 103 of
\cite{feynman1}) on the subject. Suppose we are in contact with some alien
species, but only by the exchange, say of radio signals. Any intelligible
communication needs a language and we suppose to build one doing something
similar to the one proposed in the SETI program, starting with telling aliens
what we mean by prime numbers and progressing to pictures, physics, and
chemistry information. The concept of distance may be grasped by the aliens,
e.g., by telling then how tall we are (in the mean), by expressing such number
in mutually understood wavelengths of light. They can use that information to
tell us how tall they are. We can also teach the aliens the concept of a man
lifetime by expressing such number by the number of ticks of a light-frequency
clock. To make agreement of some physical conventions and also to explain some
social procedures among men (e.g., the fact that we shake hands when we meet,
by extending our right hand) we need to explain them what is a right hand. How
to do that?

As well known, until 1957 we could not answer that question. But, after the
discover of the experiments showing parity violation in that year, we can
explain to the aliens what the right hand is by asking them to repeat the
original experiment done by Wu \cite{wu} et al with $^{60}$Co, but in such a
way that they must turn their apparatus (including the magnetic field
generator in use) until the electrons come out in the downward direction,
which we may define as the one of their local gravity pull. In such a
situation the rotating nucleus will be with their spins up, i.e., rotating in
the anti-clockwise direction as seen from the top. Before someone says that
the aliens cannot see the $^{60}$Co rotating we describe how we can teach them
to amplify this anti-clockwise rotation (as seen from the top) in order that
it becomes macroscopically visible. Indeed, all they need is to follow the
following instructions. (A) Take a spherical conductor of radius $a$ in
electrostatic equilibrium with an uniform superficial charge density with
total charge $Q$ (i.e., charge as the ones carried by the atomic nucleus of
the $^{60}$Co) and which is magnetized with its\ dipole magnetic moment \ of
modulus $\varsigma$ oriented in the same direction (the $\mathbf{\hat{z}}%
$-direction) of the magnetic moment of the $^{60}$Co nucleus in his repetition
of Wu's experiment. Such charged magnet has electromagnetic angular momentum
stored in its electromagnetic field given by $\mathbf{L}_{em}=\frac{2}%
{9}\varsigma Qa^{2}\hat{z}$\footnote{This stored angular momentum in static
electric plus magnetic field has been experimentally verified in
\cite{grala}.}. (B) The aliens are next instructed to discharge the magnet
(suspended from the roof with an insulator) through the south pole. This makes
the magnet to rotate anti-clockwise as seen from the top, in order to conserve
the total angular momentum of field plus matter. Indeed a simple calculation
\cite{sharma,gri} shows that the mechanical angular momentum acquired by the
sphere\ once completely discharged is\footnote{This result is obtained once we
neglect the magnetic field associated with the discharging current and
displacement current associated with the collapsing electric field, something
justifiable if the current is small. If the current is not small some angular
momentum will be carried by the radiation field, but of course at the end of
the discharging process the sphere will be rotating.} $\mathbf{L}_{mec}%
=\frac{2}{9}\varsigma Qa^{2}\mathbf{\hat{z}}$.

Of course, Feynman cautions us (page 107 of \cite{feynman1}) that after lots
of communication if we finally can go into space and meet the aliens
counterpart, if it happens that their leader extends its left hand to shake,
stop immediately because that is proof that he is made of antimatter. This, of
course, is because Wu's parity violation experiment constructed of antimatter
would give the opposite result.

Feynman's story is important four the objectives of this paper because it
shows that the charge argument is indeed a very week one. To fix the signal of
the charge parameters that label particles and to describe their currents, we
need to start with a convention, we need a local orientation, we need to know
what a right hand is.\footnote{More recently Elitzur and Shinitzky
\cite{elitzur} showed how to teach aliens what is right and what is left using
the space asymmetry of molecules (L and D amino acids). However, to their
method Feynman's caution also applies.}

\subsection{Electromagnetism in a medium}

Classical electromagnetism in a general medium is a very complicated subject
since admitting with Feynman that the only the fundamental physical fields $F$
and the current $J$ generated by the particles carriers must enter the game
are we immediately involved in an almost intractable many body system.
However, it seems empirical fact that the equations
\begin{align}
dF &  =0,\text{ }d\overset{}{G}\text{ }=-\overset{}{\mathbf{J}},\nonumber\\
G &  =\chi(F)\label{the pair}%
\end{align}
or
\begin{align}
dF &  =0,\text{ }d\overset{\bigtriangleup}{G}\text{ }=-\overset{\bigtriangleup
}{\mathbf{J}},\nonumber\\
\overset{\bigtriangleup}{G} &  =\overset{\triangle}{\chi}(F)\label{the impair}%
\end{align}
describes essentially all macroscopic electromagnetic phenomena on any medium
contained in a world tube in $U\subset M$.\ In those equations $\kappa$ and
$\overset{\triangle}{\kappa}$ are multiform functions \cite{femoro1,femoro2}
of the multiform variable $F$, i.e., for each $x\in U\subset M$ we have%
\begin{subequations}
\begin{align}
\left.  \chi\right\vert _{x} &  :%
%TCIMACRO{\dbigwedge \nolimits^{2}}%
%BeginExpansion
{\displaystyle\bigwedge\nolimits^{2}}
%EndExpansion
T_{x}^{\ast}M\rightarrow%
%TCIMACRO{\dbigwedge \nolimits^{2}}%
%BeginExpansion
{\displaystyle\bigwedge\nolimits^{2}}
%EndExpansion
T_{x}^{\ast}M,\label{a}\\
\left.  \overset{\triangle}{\chi}\right\vert _{x} &  :%
%TCIMACRO{\dbigwedge \nolimits^{2}}%
%BeginExpansion
{\displaystyle\bigwedge\nolimits^{2}}
%EndExpansion
T_{x}^{\ast}M\rightarrow%
%TCIMACRO{\dbigwedge \nolimits_{-}^{2}}%
%BeginExpansion
{\displaystyle\bigwedge\nolimits_{-}^{2}}
%EndExpansion
T_{x}^{\ast}M.\label{b}%
\end{align}
Consider, e.g., Eq.(\ref{b}). The multiform function $\overset{\triangle}%
{\chi}$ is phenomenologically described, e.g., using coordinates in the
\textit{ELPG} by \cite{huvan,jackson}
\end{subequations}
\begin{equation}
\overset{\triangle}{G}^{\mu\nu}=\frac{1}{2}(\overset{\triangle}{\chi}^{\mu
\nu\rho\sigma}F_{\rho\sigma}+\overset{\triangle}{\varsigma}^{\mu\nu\rho
\sigma\iota\zeta}F_{\rho\sigma}F_{\iota\zeta}+\cdots).\label{PHE}%
\end{equation}
A medium for which $\overset{\triangle}{\varsigma}^{\mu\nu\rho\sigma\iota
\zeta}\neq0$ is called nonlinear. For what follows we restrict our
considerations only to linear media. In that case the constitutive multiform
function $\overset{\triangle}{\chi}$ is a $(2,2)$-extensor field
\cite{femoro1,femoro2} and we have the decomposition \cite{hehl}
\begin{equation}
\overset{\triangle}{\chi}^{\mu\nu\rho\sigma}=\text{ }^{(1)}\overset{\triangle
}{\chi}^{\mu\nu\rho\sigma}+\text{ }^{(2)}\overset{\triangle}{\chi}^{\mu\nu
\rho\sigma}+\text{ }\overset{\triangle}{a}\text{ }^{(3)}\varepsilon^{\mu
\nu\rho\sigma},\label{chibis}%
\end{equation}
where $^{(1)}\overset{\triangle}{\chi}^{\mu\nu\rho\sigma}$is a trace free
symmetric part (with 20 independent components), $^{(2)}\overset{\triangle
}{\chi}^{\mu\nu\rho\sigma}$is the antisymmetric part (with 15 independent
components) and $\varepsilon^{\mu\nu\rho\sigma}$ is the Levi-Civita symbol
(with only 1 independent \ component). Finally, $\overset{\triangle}{a}$ is an
impair $0$-form field (also called a pseudo-scalar function) called the
\textit{axion} field.\ It has been a conjecture (called Post conjecture
\cite{post}) that $\overset{\triangle}{a}$ must be null for any medium.

However, recently it has been found that for Cr$_{2}$O$_{3}$, $\overset
{\triangle}{a}\neq0$. In \cite{hehl2} it is claimed that this fact even proves
that we must use impair forms in the description of electromagnetism. However,
those authors forget the following observation that can be found at page 22 of
de Rham's book:\medskip

\textquotedblleft Si la vari\'{e}t\'{e} $V$ est orient\'{e}, c' est-\`{a}-dire
si elle est orientable et si l"on a choisi une orientation $\varepsilon$,
\`{a} toute forme impaire $\alpha$ est associ\'{e}e une forme paire
$\varepsilon\alpha$. Par la suite, dans le cas d'une variet\'{e} orientable,
en choissant une foi pour toutes une orientation, il serai possible
d'\'{e}viter l'emploi des formes impaires. Mais pour les vari\'{e}t\'{e}s non
orientables, ce concept est r\'{e}ellement utile et naturel.\textquotedblright%
{\small \medskip}

Now, for de Rham, an orientation $\varepsilon$ is an impair $0$-form field
(i.e., an axion field) defined in the manifold $M$ (that in his book is called
$V$)\footnote{The reader can easily convince himself that this definition is
equivalent to the one given in terms of an impair and a pair volume $4$-forms
$\overset{\triangle}{\tau}_{%
%TCIMACRO{\TeXButton{g}{\sslg}}%
%BeginExpansion
\sslg
%EndExpansion
}$ and $\overset{}{\tau}_{%
%TCIMACRO{\TeXButton{g}{\sslg}}%
%BeginExpansion
\sslg
%EndExpansion
}$. Indeed, take $\varepsilon=\overset{\triangle}{\tau}_{%
%TCIMACRO{\TeXButton{g}{\sslg}}%
%BeginExpansion
\sslg
%EndExpansion
}\cdot\tau_{%
%TCIMACRO{\TeXButton{g}{\sslg}}%
%BeginExpansion
\sslg
%EndExpansion
}$.}. So, all \ that the authors of \cite{hehl2} need to do in order to have
only pair forms in their formulation of the electromagnetism of Cr$_{2}$%
O$_{3}$ is to multiply their impair objects by\ $\overset{\triangle}{a}$. So,
contrary to popular believe the existence of an axion field does not imply
that spacetime is non oriented. Quite the contrary is what is true.

For some media where $^{(2)}\overset{\triangle}{\chi}^{\mu\nu\rho\sigma}=0$,
and $\overset{\triangle}{a}$ $=0$ we even find that the constitutive extensor
may be described by \cite{post}%
\begin{equation}
\overset{\triangle}{\chi}^{\lambda\nu\sigma\kappa}=\sqrt{\det\mathbf{g}%
}(g^{\lambda\sigma}g^{\nu\kappa}-g^{\lambda\kappa}g^{\nu\sigma}) \label{cong}%
\end{equation}
where $g^{\lambda\sigma}g_{\lambda\nu}=\delta_{\nu}^{\sigma}$ and \ $g_{\mu
\nu}$ are the components of an effective metric field $\mathbf{g}=g_{\mu\nu
}d\mathrm{x}^{\mu}\otimes d\mathrm{x}^{\nu}$ for $M$. A particular medium with
such characteristic is the vacuum in the presence of a gravitational field,
but here we do not want to go deeply on this issue.

\begin{remark}
Until to this point the complete Minkowski spacetime structure $(M,%
%TCIMACRO{\TeXButton{slg}{\slg}}%
%BeginExpansion
\slg
%EndExpansion
,D,\tau_{{%
%TCIMACRO{\TeXButton{slg}{\sslg}}%
%BeginExpansion
\sslg
%EndExpansion
}},\uparrow)$ did not enter our formulation of electromagnetism. So, let us
remark, first of all that from the point of view of an experimental physicist
the parallelism rule defined by $D$ is essential, since it is this parallel
transport rule that permits him\emph{(}her\emph{)}, e.g., to make to
\textit{parallel} filamentary currents and find their interaction behavior
\emph{(}as long ago did Ampere\emph{)}.

From a mathematical point of view, the connection $D$ enter in our formulation
of electromagnetism through the introduction of the Dirac operator acting on
sections of the Clifford bundle $\mathcal{C\ell(}M,g)$.
\end{remark}

\begin{remark}
Also, for media where the constitutive extensor can be put in the form given
by \emph{Eq.(\ref{cong})} we can give an intrinsic presentation of
electromagnetism using the Clifford bundle formalism by introducing an
\textit{effective} Lorentzian spacetime $(M,\mathbf{g},\nabla,\tau
_{\mathbf{g}},\uparrow)$ where now, $\nabla$ denotes the
non-flat\emph{\footnote{This term only means that the Riemann tensor
$\mathbf{R}(\nabla)\neq0$.}} Levi-Civita connection of $\mathbf{g}$, an
effective Lorentzian metric determined by the constitutive tensor of that
effective spacetime.
\end{remark}

\begin{remark}
Before ending this section we have an important observation yet, concerning
the metric free formulation of electromagnetism as presented, e.g., in
\emph{\cite{hehl}}. There, it is admitted that $M$ is an \emph{oriented}
connected, non compact, paracompact Hausdorff space. The authors say that a
manifold with those characteristics always permits a codimension-$1$
foliation\footnote{It admits also a $1$-dimension foliation.}, a statement
that is true \emph{\cite{lawson}}. However without any additional structure we
cannot see how to foliate spacetime $M$ as time $\times$ space \emph{(}%
$\mathbb{R\times}S$\emph{)}, because we do not know a prior how to choose the
dimension that represents time. In \emph{\cite{hehl}} the authors quickly
introduce a global vector field $\mathbf{n}$ transverse to the folia, and the
$3$-dimensional manifold $S$ of the foliation is defined by a manifold
function $\sigma:M\rightarrow\mathbb{R}$ such that $\sigma(x)=$ constant and
$\mathbf{n}\lrcorner d\sigma=0$. It seems clear for us that $\mathbf{n}$ and
$\Omega=d\sigma$ are nothing more than the universal vector field and the
universal $1$-form field defining the structure of absolute space and absolute
time in the structure of Newtonian theory when that theory is formulated as a
spacetime theory \emph{(for details, see \cite{roquibo})}. Those observations
can be translated in simple words: contrary to some claims only the bare
structure of $M$ is not enough for a formulation of electromagnetic theory as
a physical theory.
\end{remark}

\section{The Clifford Bundle Formulation of Electromagnetism}

In a medium described by an effective Lorentzian spacetime $(M,\mathbf{g}%
,\nabla,\tau_{\mathbf{g}},\uparrow)$ we may present the equations of
electromagnetic theory as a single equation using the Clifford bundle
$\mathcal{C\ell(}M,\mathtt{g})$ of \textit{pair} differential forms. We recall
\cite{rodoliv2007} that in the Clifford bundle formalism the so called Dirac
operator\footnote{Please, do not confound the Dirac operator to the spin-Dirac
operator which acts on sections of a spinor bundle. See details, e.g., in
\cite{rodoliv2007}. Here we recall that in an arbitrary basis $\{e_{\mu}\}$
for $TU\subset TM$, and $\{\theta^{\mu}\}$ for $T^{\ast}U\subset T^{\ast
}M\subset\mathcal{C\ell(}M,\mathtt{g})$, the operator is given by
${\mbox{\boldmath$\partial$}:=}\theta^{\mu}\nabla_{e_{\mu}}$. Take notice that
$\mathtt{g}=g^{\mu\nu}e_{\mu}\otimes e_{\nu}$ if $\mathbf{g}=g_{\mu\nu}%
dx^{\mu}\otimes dx^{\nu}$, with $g^{\mu\nu}g_{\alpha\nu}=\delta_{\alpha}^{\mu
}$.} ${%
%TCIMACRO{\TeXButton{d}{\mbox{\boldmath$\partial$}}}%
%BeginExpansion
\mbox{\boldmath$\partial$}%
%EndExpansion
}$ acting on sections of $\mathcal{C\ell(}M,\mathtt{g})$ is given by%
\begin{equation}
{%
%TCIMACRO{\TeXButton{d}{\mbox{\boldmath$\partial$}}}%
%BeginExpansion
\mbox{\boldmath$\partial$}%
%EndExpansion
}={%
%TCIMACRO{\TeXButton{d}{\mbox{\boldmath$\partial$}}}%
%BeginExpansion
\mbox{\boldmath$\partial$}%
%EndExpansion
}\mathbf{\wedge}+{%
%TCIMACRO{\TeXButton{d}{\mbox{\boldmath$\partial$}}}%
%BeginExpansion
\mbox{\boldmath$\partial$}%
%EndExpansion
}\mathbf{\lrcorner} \label{20}%
\end{equation}
It can be shown (see, e.g., \cite{rodoliv2007}) that for a Levi-Civita
connection we have ${%
%TCIMACRO{\TeXButton{d}{\mbox{\boldmath$\partial$}}}%
%BeginExpansion
\mbox{\boldmath$\partial$}%
%EndExpansion
}\mathbf{\wedge=}d$ and \ ${%
%TCIMACRO{\TeXButton{d}{\mbox{\boldmath$\partial$}}}%
%BeginExpansion
\mbox{\boldmath$\partial$}%
%EndExpansion
}\mathbf{\lrcorner=}-$ $\delta$, where $\delta$ is the Hodge coderivative
operator, such that for any $A_{p}\in\sec%
%TCIMACRO{\dbigwedge \nolimits^{p}}%
%BeginExpansion
{\displaystyle\bigwedge\nolimits^{p}}
%EndExpansion
T^{\ast}M\hookrightarrow\mathcal{C\ell(}M,\mathtt{g})$ its action is given by:%
\begin{equation}
\delta A_{p}=(-1)^{p}\underset{\tau_{\mathbf{g}}}{\star}^{-1}d\underset
{\tau_{\mathbf{g}}}{\star}A_{p}. \label{21}%
\end{equation}
We also have that:
\begin{align}
-\delta A_{p}  &  =\mathbf{\partial\lrcorner}A_{p}=\theta^{\mu}\lrcorner
(\nabla_{e_{\mu}}A_{p}),\nonumber\\
dA_{p}  &  =\mathbf{\partial\wedge}A_{p}=\theta^{\mu}\wedge(\nabla_{e_{\mu}%
}A_{p}), \label{dd}%
\end{align}
and those expressions permit the simplification of many calculations.
Recalling that $G=\underset{\tau_{\mathbf{g}}}{\star}F$ we get that
$dG=-\mathbf{J}$ can be written defining $J=\underset{\tau_{\mathbf{g}}}%
{\star}^{-1}\mathbf{J}\in\sec%
%TCIMACRO{\dbigwedge \nolimits^{1}}%
%BeginExpansion
{\displaystyle\bigwedge\nolimits^{1}}
%EndExpansion
T^{\ast}M\hookrightarrow\mathcal{C\ell(}M,\mathtt{g})$ as $\delta F=-J$.
Indeed, we have,
\begin{align*}
d\underset{\tau_{\mathbf{g}}}{\star}F  &  =-\mathbf{J,}\\
\underset{\tau_{\mathbf{g}}}{\star}^{-1}d\underset{\tau_{\mathbf{g}}}{\star}F
&  =-\underset{\tau_{\mathbf{g}}}{\star}^{-1}\mathbf{J,}\\
\delta F  &  =-J
\end{align*}
Then, the two equations $dF=0$ and $\delta F=-J$ can be \textit{summed} if we
suppose (as it is licit to do in the Clifford bundle $\mathcal{C\ell
(}M,\mathtt{g})$) that $F,G\in\sec%
%TCIMACRO{\dbigwedge \nolimits^{2}}%
%BeginExpansion
{\displaystyle\bigwedge\nolimits^{2}}
%EndExpansion
T$ $^{\ast}M\hookrightarrow\mathcal{C\ell(}M,\mathtt{g})$ and $\mathbf{J}%
\in\sec%
%TCIMACRO{\dbigwedge \nolimits^{3}}%
%BeginExpansion
{\displaystyle\bigwedge\nolimits^{3}}
%EndExpansion
T^{\ast}M\hookrightarrow\mathcal{C\ell(}M,\mathtt{g})$ and we get
\textit{Maxwell equation\footnote{No misprint here! Parodying Thirring
\cite{thirring} that said that the equations $dF=0$ and $\delta F=0$ were the
20$^{th}$ century presentation of Maxwell equations, we say that the single
equation ${%
%TCIMACRO{\TeXButton{d}{\mbox{\boldmath$\partial$}}}%
%BeginExpansion
\mbox{\boldmath$\partial$}%
%EndExpansion
}F=J$ is the 21$^{th}$ century presentation of Maxwell equations.}}%
\begin{equation}
{%
%TCIMACRO{\TeXButton{d}{\mbox{\boldmath$\partial$}}}%
%BeginExpansion
\mbox{\boldmath$\partial$}%
%EndExpansion
}F=J. \label{22}%
\end{equation}

\begin{remark}
We now show that \emph{Eq.(\ref{22})} can also be obtained directly from the
de Rham formulation of electromagnetism that uses pair and impair forms.
Indeed, all that is need is to verify that the formula $d\overset
{\bigtriangleup}{G}$ $=-\overset{\bigtriangleup}{\mathbf{J}}$ in
\emph{Eq.(\ref{19})} --- where $\overset{\bigtriangleup}{G}$ $=\underset
{\overset{\bigtriangleup}{\tau_{_{\mathbf{g}}}}}{\star}F$ --- can be written
as $\delta F=-J$. Indeed, we have
\begin{align*}
d\overset{\bigtriangleup}{G}\text{ }  &  =-\overset{\bigtriangleup}%
{\mathbf{J}},\\
d\underset{\overset{\bigtriangleup}{\tau_{_{\mathbf{g}}}}}{\star}F  &
=-\overset{\bigtriangleup}{\mathbf{J}},\\
\underset{\overset{\bigtriangleup}{\tau_{\mathbf{g}}}}{\star}^{-1}%
d\underset{\overset{\bigtriangleup}{\tau_{\mathbf{g}}}}{\star}F  &
=-\underset{\overset{\bigtriangleup}{\tau_{\mathbf{g}}}}{\star}^{-1}%
\overset{\bigtriangleup}{\mathbf{J}},
\end{align*}
and it is trivial to verify that $\underset{\overset{\bigtriangleup}%
{\tau_{\mathbf{g}}}}{\star}^{-1}d\underset{\overset{\bigtriangleup}%
{\tau_{\mathbf{g}}}}{\star}F=\delta F$ and \ $\underset{\overset
{\bigtriangleup}{\tau_{\mathbf{g}}}}{\star}^{-1}\overset{\bigtriangleup
}{\mathbf{J}}$ $=\underset{\tau_{\mathbf{g}}}{\star}^{-1}\mathbf{J=}$
$J\in\sec%
%TCIMACRO{\dbigwedge \nolimits^{1}}%
%BeginExpansion
{\displaystyle\bigwedge\nolimits^{1}}
%EndExpansion
T$ $^{\ast}M$. We conclude that the Clifford bundle formulation of
electromagnetism given by Maxwell equation (\emph{Eq.(\ref{19})}) is general
enough to permit the two formulations of electromagnetism given above.
\end{remark}

We are now ready to complete the formulation of electrodynamics as a physical
theory. We restrict our presentation here in the case where the \ existence of
the gravitational field must be ignored. As such our formulation will use the
Minkowski spacetime structure introduced above and the Clifford bundle
$\mathcal{C\ell(}M,g).$

\section{The Energy-momentum 1-forms of the Electromagnetic and the Matter
Fields}

We start from Maxwell equation (with $J$ the current of charged particles
introduced by Eq.(\ref{curre1}))%

\begin{equation}
{%
%TCIMACRO{\TeXButton{d}{\mbox{\boldmath$\partial$}}}%
%BeginExpansion
\mbox{\boldmath$\partial$}%
%EndExpansion
}F=J \label{ma}%
\end{equation}
where in what follows, ${%
%TCIMACRO{\TeXButton{d}{\mbox{\boldmath$\partial$}}}%
%BeginExpansion
\mbox{\boldmath$\partial$}%
%EndExpansion
=}\theta^{\alpha}D_{e_{\alpha}}=\gamma^{\mu}\frac{\partial}{\partial
\mathrm{x}^{\mu}}$is the Dirac operator written with a general pair of dual
basis $\{e_{\alpha}\}$ and $\{\theta^{\alpha}\}$ for $TU\subset$ $TM$ and
$T^{\ast}U\subset T^{\ast}M$ and with the basis $\{\frac{\partial}%
{\partial\mathrm{x}^{\mu}}\}$ and $\{\gamma^{\mu}=d\mathrm{x}^{\mu}\}$ for
$TM$ and $T^{\star}M$, with $\{\mathrm{x}^{\mu}\}$ coordinates in the
Einstein-Lorentz-Poincar\'{e} gauge. Given Eq.(\ref{ma}) its reverse is
\begin{equation}
\tilde{F}\overleftarrow{{%
%TCIMACRO{\TeXButton{d}{\mbox{\boldmath$\partial$}}}%
%BeginExpansion
\mbox{\boldmath$\partial$}%
%EndExpansion
}}=J \label{mar}%
\end{equation}
where $\tilde{F}\overleftarrow{{%
%TCIMACRO{\TeXButton{d}{\mbox{\boldmath$\partial$}}}%
%BeginExpansion
\mbox{\boldmath$\partial$}%
%EndExpansion
}}:=(D_{e_{\alpha}}\tilde{F})\theta^{\alpha}=(\frac{\partial}{\partial
\mathrm{x}^{\mu}}\tilde{F})\gamma^{\mu}$. Multiplying Eq.(\ref{ma}) on the
left by $\tilde{F}$ and Eq.(\ref{mar}) on the right by $F$ and summing the
resulting equations we get%
\begin{equation}
\frac{1}{2}[\tilde{F}({%
%TCIMACRO{\TeXButton{d}{\mbox{\boldmath$\partial$}}}%
%BeginExpansion
\mbox{\boldmath$\partial$}%
%EndExpansion
}F)+(\tilde{F}\overleftarrow{{%
%TCIMACRO{\TeXButton{d}{\mbox{\boldmath$\partial$}}}%
%BeginExpansion
\mbox{\boldmath$\partial$}%
%EndExpansion
}})F]=\frac{1}{2}(\tilde{F}J+JF), \label{int}%
\end{equation}
Now, let $n=n^{\alpha}\gamma_{\alpha}\in\sec%
%TCIMACRO{\dbigwedge \nolimits^{1}}%
%BeginExpansion
{\displaystyle\bigwedge\nolimits^{1}}
%EndExpansion
T^{\ast}M\hookrightarrow\mathcal{C\ell(}M,g)$ and ${%
%TCIMACRO{\TeXButton{d}{\mbox{\boldmath$\partial$}}}%
%BeginExpansion
\mbox{\boldmath$\partial$}%
%EndExpansion
}_{n}=\gamma^{\alpha}\frac{\partial}{\partial n^{\alpha}}$ acting on multiform
functions of the multiform variable $n$. Consider moreover the extensor field
\footnote{An extensor field $T:%
%TCIMACRO{\dbigwedge \nolimits^{1}}%
%BeginExpansion
{\displaystyle\bigwedge\nolimits^{1}}
%EndExpansion
T^{\ast}M\rightarrow%
%TCIMACRO{\dbigwedge \nolimits^{1}}%
%BeginExpansion
{\displaystyle\bigwedge\nolimits^{1}}
%EndExpansion
T^{\ast}M$, $n\mapsto T(n)$ is a linear multiform function of \ the form field
$n$.} $T(n)=\frac{1}{2}\tilde{F}nF$. Now, observe that if we apply
$\gamma^{\alpha}\cdot{%
%TCIMACRO{\TeXButton{d}{\mbox{\boldmath$\partial$}}}%
%BeginExpansion
\mbox{\boldmath$\partial$}%
%EndExpansion
}_{n}$ to the multiform function $\mathbf{f}(n)=\frac{\partial n}%
{\partial\mathrm{x}^{\alpha}}$ we get\footnote{See details on the derivation
of multiform functions in \cite{mofero}.}
\begin{align}
\gamma^{\alpha}\cdot{%
%TCIMACRO{\TeXButton{d}{\mbox{\boldmath$\partial$}}}%
%BeginExpansion
\mbox{\boldmath$\partial$}%
%EndExpansion
}_{n}\frac{\partial n}{\partial\mathrm{x}^{\alpha}}  &  =\eta^{\alpha\mu}%
\frac{\partial}{\partial n^{\mu}}(\frac{\partial}{\partial\mathrm{x}^{\alpha}%
}n^{\beta}\gamma_{\beta})\nonumber\\
&  =\eta^{\alpha\mu}\frac{\partial}{\partial\mathrm{x}^{\alpha}}(\delta_{\mu
}^{\beta}\gamma_{\beta})=0. \label{ide}%
\end{align}
Using Eq.(\ref{ide}) we can write the first member of Eq.(\ref{int}) as
\begin{align}
&  \tilde{F}\gamma^{\lambda}\frac{\partial F}{\partial\mathrm{x}^{\lambda}%
}+\frac{\partial\tilde{F}}{\partial\mathrm{x}^{\lambda}}\gamma^{\lambda
}F\nonumber\\
&  =\gamma^{\lambda}\cdot{%
%TCIMACRO{\TeXButton{d}{\mbox{\boldmath$\partial$}}}%
%BeginExpansion
\mbox{\boldmath$\partial$}%
%EndExpansion
}_{n}\left(  \tilde{F}n\frac{\partial F}{\partial\mathrm{x}^{\lambda}}%
+\tilde{F}\frac{\partial n}{\partial\mathrm{x}^{\lambda}}F+\frac
{\partial\tilde{F}}{\partial\mathrm{x}^{\lambda}}nF\right) \nonumber\\
&  =\gamma^{\lambda}\cdot{%
%TCIMACRO{\TeXButton{d}{\mbox{\boldmath$\partial$}}}%
%BeginExpansion
\mbox{\boldmath$\partial$}%
%EndExpansion
}_{n}\frac{\partial}{\partial\mathrm{x}^{\lambda}}(\tilde{F}nF)\label{xx}\\
&  =\frac{\partial}{\partial\mathrm{x}^{\lambda}}(\tilde{F}\gamma^{\lambda
}F)\nonumber
\end{align}
On the other hand the second member of Eq.(\ref{int}) is just $-J\lrcorner F$.
So, we have
\begin{equation}
\frac{\partial}{\partial\mathrm{x}^{\alpha}}T^{\alpha}=J\lrcorner F, \label{q}%
\end{equation}
with the $T^{\alpha}\in\sec%
%TCIMACRO{\dbigwedge \nolimits^{1}}%
%BeginExpansion
{\displaystyle\bigwedge\nolimits^{1}}
%EndExpansion
T^{\ast}M\hookrightarrow\mathcal{C\ell(}M,g)$ given by\footnote{An equation
equivalent to Eq.(\ref{emt}) has been discovered by M. Riesz \cite{riesz}.} \
\begin{equation}
T^{\alpha}=\frac{1}{2}F\gamma^{\alpha}\tilde{F} \label{emt}%
\end{equation}
being the pair energy-momentum $1$-forms of the electromagnetic field. Indeed,
a simple calculation shows that
\begin{equation}
T^{\alpha\beta}=T^{\alpha}\cdot\gamma^{\beta}=\eta^{\alpha\mu}F_{\mu\lambda
}F^{\lambda\beta}+\frac{1}{4}\eta^{\alpha\beta}F_{\mu\nu}F^{\mu\nu},
\label{cemt}%
\end{equation}
a well known formula. Now, we contract Eq.(\ref{q}) \ on the left with
$\gamma^{\alpha}$ getting%
\begin{equation}
{%
%TCIMACRO{\TeXButton{d}{\mbox{\boldmath$\partial$}}}%
%BeginExpansion
\mbox{\boldmath$\partial$}%
%EndExpansion
\lrcorner}T^{\alpha}=\gamma^{\alpha}\lrcorner(J\lrcorner F) \label{xxx}%
\end{equation}
Now\footnote{The sequence of identities in Eq.(\ref{ident}) may be found in
Section 2.4.2 of \cite{rodoliv2007}.},%
\begin{align}
\gamma^{\alpha}\lrcorner(J\lrcorner F)  &  =(\gamma^{\alpha}\wedge J)\lrcorner
F=-(\gamma^{\alpha}\wedge J)\cdot F\nonumber\\
&  =-F\cdot(\gamma^{\alpha}\wedge J)=-(\gamma^{\alpha}\lrcorner F)\cdot J,
\label{ident}%
\end{align}
and Eq.(\ref{xxx}) becomes (taking into account that ${%
%TCIMACRO{\TeXButton{d}{\mbox{\boldmath$\partial$}}}%
%BeginExpansion
\mbox{\boldmath$\partial$}%
%EndExpansion
\lrcorner}T^{\alpha}=-\delta T^{\alpha}$)
\begin{equation}
\delta T^{\alpha}=(\gamma^{\alpha}\lrcorner F)\cdot J \label{l}%
\end{equation}
Defining $f^{\alpha}\in\sec%
%TCIMACRO{\dbigwedge \nolimits^{4}}%
%BeginExpansion
{\displaystyle\bigwedge\nolimits^{4}}
%EndExpansion
T^{\ast}M$ as the \textit{pair }density of force by
\begin{equation}
f^{\alpha}=[(\gamma^{\alpha}\lrcorner F)\cdot J]\tau_{g}, \label{f0}%
\end{equation}
where $\tau_{g}$ is a pair metric volume element, we obtain (the equivalent expressions)%

\begin{align}
f^{\alpha}  &  =[(\gamma^{\alpha}\lrcorner F)\cdot J)]\tau_{g}=\underset
{\tau_{\mathbf{%
%TCIMACRO{\TeXButton{g}{\sslg}}%
%BeginExpansion
\sslg
%EndExpansion
}}}{\star}[(\gamma^{\alpha}\lrcorner F)\lrcorner J]=(\gamma^{\alpha}\lrcorner
F)\wedge\underset{\tau_{\mathbf{%
%TCIMACRO{\TeXButton{g}{\sslg}}%
%BeginExpansion
\sslg
%EndExpansion
}}}{\star}J\nonumber\\
&  =(\gamma^{\alpha}\lrcorner F)\wedge\mathbf{J}, \label{f}%
\end{align}
where, in particular, the last one is the pair density of force.

\begin{remark}
Note that we could return to Eq.(\ref{10}) and get an impair density of force
simply by replacing the pair volume element $\tau_{g}$ by an impair volume
element $\overset{\triangle}{\tau}_{g}$, i.e., defining
\begin{equation}
\overset{\triangle}{f}\text{ }^{\alpha}:=[(\gamma^{\alpha}\lrcorner F)\cdot
J]\text{ }\overset{\triangle}{\tau}_{\mathbf{%
%TCIMACRO{\TeXButton{g}{\sslg}}%
%BeginExpansion
\sslg
%EndExpansion
}}. \label{f01}%
\end{equation}
Such a formula was \textit{postulated} in the presentation of electromagnetism
in \cite{hehl} However, as we just saw, that postulate for the coupling of the
field $F$ with the current $\mathbf{J}$ is not necessary in our approach,
since that force density is already contained in Maxwell equation
(Eq.(\ref{ma})). We now write Eq.(\ref{l}) as
\begin{equation}
d\underset{\tau_{\mathbf{%
%TCIMACRO{\TeXButton{g}{\sslg}}%
%BeginExpansion
\sslg
%EndExpansion
}}}{\star}T^{\alpha}=(\gamma^{\alpha}\lrcorner F)\wedge\underset
{\tau_{\mathbf{%
%TCIMACRO{\TeXButton{g}{\sslg}}%
%BeginExpansion
\sslg
%EndExpansion
}}}{\star}J, \label{ft}%
\end{equation}
or
\begin{equation}
d\underset{\overset{\triangle}{\tau}_{\mathbf{%
%TCIMACRO{\TeXButton{g}{\sslg}}%
%BeginExpansion
\sslg
%EndExpansion
}}}{\star}T^{\alpha}=(\gamma^{\alpha}\lrcorner F)\wedge\underset
{\overset{\triangle}{\tau}_{\mathbf{%
%TCIMACRO{\TeXButton{g}{\sslg}}%
%BeginExpansion
\sslg
%EndExpansion
}}}{\star}J \label{ft'}%
\end{equation}
and both, of course, in components reads%
\begin{equation}
\partial_{\nu}T^{\alpha\nu}=J_{\nu}F^{\nu\alpha} \label{ftb}%
\end{equation}

\end{remark}

\subsection{Total Energy-Momentum Conservation and the Knockdown of the Charge
Argument}

Eq.(\ref{ftb}) asserts that the energy momentum tensor of the electromagnetic
field is not conserved. We expect that the \textit{total} energy-momentum of
the field and the charged particles is \textit{conserved} since there is not a
single experiment in Physics contradicting it, and so without much ado,
recalling the definition of the \textit{pair} energy-momentum 1-forms of the
charged matter, the $\mathbf{T}^{\alpha}$ given by Eq.(\ref{emtm}) we
postulated that:%
\begin{equation}
\delta T^{\alpha}+\delta\mathbf{T}^{\alpha}=0, \label{temc}%
\end{equation}
which may be written as:%
\begin{equation}
(\gamma^{\alpha}\lrcorner F)\wedge\underset{\tau_{\mathbf{%
%TCIMACRO{\TeXButton{g}{\sslg}}%
%BeginExpansion
\sslg
%EndExpansion
}}}{\star}J=-d\underset{\tau_{\mathbf{%
%TCIMACRO{\TeXButton{g}{\sslg}}%
%BeginExpansion
\sslg
%EndExpansion
}}}{\star}\mathbf{T}^{\alpha}. \label{ftt}%
\end{equation}
or%
\begin{equation}
(\gamma^{\alpha}\lrcorner F)\wedge\underset{\overset{\triangle}{\tau
}_{\mathbf{%
%TCIMACRO{\TeXButton{g}{\sslg}}%
%BeginExpansion
\sslg
%EndExpansion
}}}{\star}J=-d\underset{\overset{\triangle}{\tau}_{\mathbf{%
%TCIMACRO{\TeXButton{g}{\sslg}}%
%BeginExpansion
\sslg
%EndExpansion
}}}{\star}\mathbf{T}^{\alpha}. \label{ftth}%
\end{equation}

Eq.(\ref{ftt}) is of course the \textit{Lorentz force law}, and all objects in
it\textit{ }are pair forms. On the other hand Eq.(\ref{ftth}) is also an
expression of the Lorentz force law and there are objects on it that are pair
and others that are impair forms. Both equations give in our opinion the
correct description of physical phenomena. However let us analyze here
Eq.(\ref{ftt}), since it permits us to knockdown again the \textit{charge
argument} \cite{frankel,post, post1,post2, kiehn}\emph{ }\ which says that the
density of current must be an impair $3$-form, i.e., that we must use
$\overset{\triangle}{\mathbf{J}}=\underset{\overset{\triangle}{\tau}_{\mathbf{%
%TCIMACRO{\TeXButton{g}{\sslg}}%
%BeginExpansion
\sslg
%EndExpansion
}}}{\star}J\in\sec%
%TCIMACRO{\dbigwedge \nolimits^{3}}%
%BeginExpansion
{\displaystyle\bigwedge\nolimits^{3}}
%EndExpansion
T^{\ast}M$ to calculate without ambiguity the charge in a certain region, say
$U\in M$. To see this, recall that the total energy-momentum 1-form of matter
in $U$ at time \textrm{x}$^{0}=t$ in an inertial reference frame
$\mathbf{I}=\frac{\partial}{\partial\mathrm{x}^{0}}$ is
\begin{equation}
P(t)=\left(
%TCIMACRO{\dint \nolimits_{U}}%
%BeginExpansion
{\displaystyle\int\nolimits_{U}}
%EndExpansion
\underset{\tau_{\mathbf{%
%TCIMACRO{\TeXButton{g}{\sslg}}%
%BeginExpansion
\sslg
%EndExpansion
}}}{\star}\mathbf{T}^{\alpha}\right)  \gamma_{\alpha}. \label{p}%
\end{equation}
Now, if we change the orientation of $U$ two things happen. What was called
electric charge $q^{(i)}=%
%TCIMACRO{\dint \nolimits_{U}}%
%BeginExpansion
{\displaystyle\int\nolimits_{U}}
%EndExpansion
\left.  \underset{\tau_{\mathbf{%
%TCIMACRO{\TeXButton{g}{\sslg}}%
%BeginExpansion
\sslg
%EndExpansion
}}}{\star}J^{(i)}\right\vert _{\sigma^{(i)}}$ of the $i$-particle changes into
$-q^{(i)}$. $F$ changes into $-F$ (despite the fact that it is a pair form)
because of the formula used to calculate it (see Appendix). If we are
interested in the motion of only a single small particle modeled by a thin
world tube in an external field $F$ , when integrating Eq.(\ref{p}) we get
that what we originally called energy at time $t$, $E^{(i)}(t)=%
%TCIMACRO{\dint \nolimits_{U}}%
%BeginExpansion
{\displaystyle\int\nolimits_{U}}
%EndExpansion
\left.  \underset{\tau_{\mathbf{%
%TCIMACRO{\TeXButton{g}{\sslg}}%
%BeginExpansion
\sslg
%EndExpansion
}}}{\star}\mathbf{T}^{0}\right\vert _{\sigma^{(i)}}$ of $i$-particle changes
into $-E^{(i)}(t)$. This sign changes in Eq.(\ref{p}) is compensated by the
sign change that occurs in $%
%TCIMACRO{\dint \nolimits_{U}}%
%BeginExpansion
{\displaystyle\int\nolimits_{U}}
%EndExpansion
(\gamma^{\alpha}\lrcorner F)\wedge\underset{\overset{\triangle}{\tau
}_{\mathbf{%
%TCIMACRO{\TeXButton{g}{\sslg}}%
%BeginExpansion
\sslg
%EndExpansion
}}}{\star}J_{(i)}$ and it follows that the prediction for the trajectory of
that particle does not change if we change the orientation of $U$. And since
trajectories of particles are all what are experimentally detected, it follows
that the formulation of electrodynamics with only pair forms is compatible
with the experimental facts.

\section{The Engineering\ Formulation of Electromagnetism Without Axial Vector
Fields}

We recall (see details in \cite{rodoliv2007}) that for any $x\in M$,
${\mathcal{C}\ell}(T_{x}^{\ast}M,g_{x})\simeq\mathbb{R}_{1,3}\simeq
\mathbb{H}(2)$, is the so called spacetime algebra. The even elements of
$\mathbb{R}_{1,3}$ close a subalgebra called the Pauli algebra. That even
subalgebra is denoted by $\mathbb{R}_{1,3}^{0}\simeq\mathbb{R}_{3,0}%
\simeq\mathbb{C}(2)$. Also, $\mathbb{H}(2)$ is the algebra of the $2\times2$
quaternionic matrices and $\mathbb{C}(2)$ is the algebra of the $2\times2$
complex matrices. \ There is an isomorphism $\mathbb{R}_{1,3}^{0}%
\simeq\mathbb{R}_{3,0}$ as the reader can easily convince himself. Choose a
global orthonormal tetrad coframe $\{\gamma^{\mu}\}$, $\gamma^{\mu
}=d\mathrm{x}^{\mu}$, \ $\mu=0,1,2,3$, and let $\{\gamma_{\mu}\}$ be the
reciprocal tetrad of $\{\gamma^{\mu}\}$, i.e., $\gamma_{\nu}\cdot\gamma^{\mu
}=\delta_{\nu}^{\mu}$ . Now, put
\begin{equation}
\sigma_{i}=\gamma_{i}\gamma_{0},\text{ }\mathbf{i}=-\gamma^{0}\gamma^{1}%
\gamma^{2}\gamma^{3}=-\gamma^{5}. \label{mp1}%
\end{equation}

Observe that $\mathbf{i}$ commutes with bivectors and thus \textit{acts} like
the imaginary unity $\mathrm{i}=\sqrt{-1}$ in the even subbundle
${\mathcal{C}\ell}^{0}(M,g)=\bigcup\nolimits_{x\in M}{\mathcal{C}\ell}%
^{0}(T_{x}^{\ast}M,g_{x})\hookrightarrow{\mathcal{C}\ell}(M,\mathtt{g})$,
which may be called the \textit{Pauli bundle. }Now, the electromagnetic field
is represented in ${\mathcal{C}\ell}(M,g)$ by $F={\frac{1}{2}}F{^{\mu\nu
}\gamma_{\mu}\wedge\gamma_{\nu}}\in\sec{\bigwedge}^{2}T^{\ast}M\hookrightarrow
\sec{\mathcal{C}\ell}(M,g)$ with%

\begin{equation}
F^{\mu\nu}=\left(
\begin{array}
[c]{cccc}%
0 & -E_{1} & -E_{2} & -E_{3}\\
E_{1} & 0 & -B_{3} & B_{2}\\
E_{2} & B_{3} & 0 & -B_{1}\\
E_{3} & -B_{2} & B_{1} & 0
\end{array}
\right)  \;, \label{faraday contra}%
\end{equation}
where ($E_{1},E_{2},E_{3}$) and ($B_{1},B_{2},B_{3}$) are the \textit{usual}
Cartesian components of the electric and magnetic fields. Then, as it is easy
to verify we can write%

\begin{equation}
F=\mathbf{E}+\mathbf{iB},
\end{equation}
with, $\mathbf{E}=\sum\nolimits_{i=1}^{3}E_{i}\sigma_{i}$, $\mathbf{B}%
=\sum\nolimits_{i=1}^{3}B_{i}\sigma_{i}$.

\begin{remark}
Although $\mathbf{E}$ and $\mathbf{B}$ are $2$-form fields in ${\mathcal{C}%
\ell}(M,\mathtt{g})$ they may be identified, once we fix an inertial reference
frame \emph{(i}.e., fix the $\gamma^{0}$ field\emph{)} with time dependent
Euclidean vector fields $\vec{E},\vec{B}$ and thus we call them
\textquotedblleft\ Euclidean vector fields\textquotedblright\ in
${\mathcal{C}\ell}^{0}(M,\mathtt{g}).$
\end{remark}

For the electric current density $J_{e}=\rho\gamma^{0}+J^{i}\gamma_{i}$ we can
write
\begin{equation}
\gamma_{0}J_{e}=\rho-\mathbf{j}=\rho-J^{i}\sigma_{i}.
\end{equation}
For the Dirac operator we have
\begin{equation}
\gamma_{0}{%
%TCIMACRO{\TeXButton{d}{\mbox{\boldmath$\partial$}}}%
%BeginExpansion
\mbox{\boldmath$\partial$}%
%EndExpansion
}=\frac{\partial}{\partial x^{0}}+\sum\limits_{i=1}^{3}\sigma_{i}\partial
_{i}=\frac{\partial}{\partial t}+\nabla.
\end{equation}
Multiplying both members of Eq.(\ref{22}) on the left by $\gamma_{0}$ we
obtain
\begin{align}
\gamma_{0}{%
%TCIMACRO{\TeXButton{d}{\mbox{\boldmath$\partial$}}}%
%BeginExpansion
\mbox{\boldmath$\partial$}%
%EndExpansion
}F  &  =\gamma_{0}J,\nonumber\\
(\frac{\partial}{\partial t}+\nabla)(\mathbf{E}+\mathbf{iB})  &
=\rho-\mathbf{j} \label{mp2}%
\end{align}
From Eq.(\ref{mp2}) we obtain
\begin{equation}
\partial_{0}\mathbf{E}+\mathbf{i}\partial_{0}\mathbf{B}+\nabla\bullet
\mathbf{E}+\nabla\curlywedge\mathbf{E}+\mathbf{i}\nabla\bullet\mathbf{B}%
+\mathbf{i}\nabla\curlywedge\mathbf{B}=\rho-\mathbf{j}. \label{mp3}%
\end{equation}
In Eq.(\ref{mp3}) for any \textquotedblleft vector field\textquotedblright%
\ $\mathbf{A}\in$ ${\mathcal{C}\ell}^{0}(M,\mathtt{g})(\hookrightarrow
{\mathcal{C}\ell}(M,\mathtt{g}))$,%

\begin{align}
\nabla\bullet\mathbf{A}  &  =\sigma_{i}\bullet(\partial_{i}\mathbf{A}%
),\nonumber\\
\nabla\curlywedge\mathbf{A}  &  =\sigma_{i}\curlywedge(\partial_{i}%
\mathbf{A}), \label{mp4}%
\end{align}
with the symbols $\bullet$ being defined through%
\begin{align}
\sigma_{i}\bullet\sigma_{j}  &  =\frac{1}{2}(\sigma_{i}\sigma_{j}+\sigma
_{j}\sigma_{i})=\delta_{ij},\nonumber\\
\sigma_{i}\curlywedge\sigma_{j}  &  =\frac{1}{2}(\sigma_{i}\sigma_{j}%
-\sigma_{j}\sigma_{i}). \label{mp5}%
\end{align}

We define next the \emph{vector} product of two ``vector
fields\textquotedblright\ $\mathbf{C}$ $=\sum\nolimits_{i=1}^{3}C_{i}%
\sigma_{i}$ and $\mathbf{D}=\sum\nolimits_{i=1}^{3}D_{i}\sigma_{i}$ as the
dual (see, e.g., \cite{rodoliv2007}) \ of the \textquotedblleft bivector
field\textquotedblright\ $\mathbf{C}$ $\curlywedge\mathbf{D}$ through the
formula%
\begin{equation}
\mathbf{C}\times\mathbf{D}=-\mathbf{i(C\curlywedge D)}. \label{vector product}%
\end{equation}

Finally, for any \textquotedblleft vector field\textquotedblright%
\ $\mathbf{A}\in$ ${\mathcal{C}\ell}^{0}(M,\mathtt{g})(\hookrightarrow
{\mathcal{C}\ell}(M,\mathtt{g}))$ we define the \textit{rotational operator}
$\nabla\times$ by
\begin{equation}
\nabla\times\mathbf{A}=-\mathbf{i(}\nabla\curlywedge\mathbf{A)}.
\label{rotational}%
\end{equation}

Using these concepts we obtain from Eq.(\ref{mp3})\emph{ }by equating terms
with the same grades (in the Pauli subbundle)%

\begin{equation}%
\begin{tabular}
[c]{ccccc}%
(a) & $\nabla\cdot\mathbf{E}=\rho,$ &  & (b) & $\nabla\times\mathbf{B}%
-\partial_{0}\mathbf{E}=\mathbf{j},$\\
(c) & $\nabla\times\mathbf{E}+\partial_{0}\mathbf{B}=0,$ &  & (d) &
$\nabla\cdot\mathbf{B}=0,$%
\end{tabular}
\ \ \ \ \ \label{vector maxwell}%
\end{equation}
which we recognize as the system of Maxwell equations in the usual vector
(engineering) notation. However, the following remark is necessary.

\begin{remark}
From the above developments we see that a direct formulation of
electromagnetism using time dependent fields, which are taken as sections of
the Pauli subbundle ${\mathcal{C}\ell}^{0}(M,\mathtt{g})$, uses \emph{only}
vector fields, once an \emph{orientation (}say $\mathbf{i}$\emph{)} is fixed,
thus dispensing the axial vector fields of the traditional Gibbs-Heaviside
formulation and the more sophisticated formalism of tensors and tensor
densities introduced by \emph{\cite{scho}} and presented as a necessity by
some other authors. Moreover, \emph{Eq.(\ref{vector product})} leaves also
clear that the definition of the vector product depends --- in each inertial
frame \emph{(}i.e., when we fix field $\gamma^{0}$\emph{)} --- on the choice
of an orientation in the affine Euclidean rest space $S$ \emph{\cite{sawu}} of
that frame. It implies that if we change the orientation of $S$, i.e., choose
$-\mathbf{i}$ \ \emph{(}instead of $\mathbf{i}$) in the definition of the
vector product, we need to change $\mathbf{B}\mapsto-\mathbf{B}$, which means
that the circulation of the magnetic field around a \emph{(}very long\emph{)}
wire conducting current is conventional \emph{\cite{frankel}}.
\end{remark}

\section{Conclusions}

We showed that in any relativistic spacetime $(M,\mathbf{g},D,\tau
_{\mathbf{g}},\uparrow)$, which is necessarily an \textit{orientable} and
\textit{time orientable} manifold, electromagnetism can be coherently
formulated using only \textit{pair} form fields or \textit{pair} and
\textit{impair} form fields, contrary to some claims appearing in the
literature. The use of pair and impair form fields is necessary only if a non
orientable manifold models our universe. However, a manifold of this kind
cannot (according to a well known result \cite{geroch}) represent the
spacetime of our universe, where spinor fields live. Moreover we showed that
using the Clifford bundle of (pair) forms we can give a formulation of
electromagnetism that is compatible with those two formulations using only
pair form fields or pair and impair form fields. Each one of those
formulations depends only on a mathematical choice that does not seem to imply
in any observable consequence.

An eventual objection to our formulation not discussed above, appeared in
\cite{gsponer}, which claims that the description of electromagnetism using
the Clifford bundle formalism is not consistent if magnetic monopoles exist.
Now, using that formalism the generalized Maxwell equations read
\begin{equation}
{%
%TCIMACRO{\TeXButton{d}{\mbox{\boldmath$\partial$}}}%
%BeginExpansion
\mbox{\boldmath$\partial$}%
%EndExpansion
}F=J-\text{ }\underset{\tau_{\mathbf{%
%TCIMACRO{\TeXButton{g}{\sslg}}%
%BeginExpansion
\sslg
%EndExpansion
}}}{\star}J_{m}, \label{gm}%
\end{equation}
where $J$ is the pair electric current $1$-form field and $J_{m}$ is the pair
magnetic current $1$-form \ field, and the claim in \cite{gsponer} is that the
Clifford bundle formalism implies that $J_{m}=0$ \ Since this statement
appears from time to time it is opportune to recall here that it has been
proved wrong in \cite{rod, rodoliv2007}, since based on a misunderstanding
that says that if the electric charges are scalars the magnetic charges must
be pseudo-scalars. It also must be said that even if magnetic monopoles do not
exist, Eq.(\ref{gm}) is important. The reason is the following. It can be
shown that in the Clifford bundle formalism the standard Dirac equation
describing, say the interaction of an electron field with the electromagnetic
field, is represented by an equation called the the Dirac-Hestenes
\cite{hestenes} equation\ which can be putted in that form
\begin{equation}
{%
%TCIMACRO{\TeXButton{d}{\mbox{\boldmath$\partial$}}}%
%BeginExpansion
\mbox{\boldmath$\partial$}%
%EndExpansion
\psi}\text{ }{\gamma}^{2}\gamma^{1}+m\psi\gamma^{0}+qA\psi=0, \label{dirac}%
\end{equation}
where $\psi$ is a Dirac-Hestenes spinor field \cite{mr,rm,rodoliv2007}, a
mathematical object represented in a given inertial frame $\mathbf{I}%
=\partial/\partial\mathrm{x}^{0}$ \ and once a spin-frame is fixed by a non
homogeneous even section of the Clifford bundle\footnote{More precisely,
Dirac-Hestenes spinor fields are some equivalent classes of even \textit{non
homogeneous} differential forms. See, \cite{mr,rm} for details.},%
\begin{equation}
\psi=S+F+\tau_{\mathbf{%
%TCIMACRO{\TeXButton{g}{\sslg}}%
%BeginExpansion
\sslg
%EndExpansion
}}P\in\sec(%
%TCIMACRO{\dbigwedge \nolimits^{0}}%
%BeginExpansion
{\displaystyle\bigwedge\nolimits^{0}}
%EndExpansion
T^{\ast}M+%
%TCIMACRO{\dbigwedge \nolimits^{2}}%
%BeginExpansion
{\displaystyle\bigwedge\nolimits^{2}}
%EndExpansion
T^{\ast}M+%
%TCIMACRO{\dbigwedge \nolimits^{4}}%
%BeginExpansion
{\displaystyle\bigwedge\nolimits^{4}}
%EndExpansion
T^{\ast}M).
\end{equation}
It can then easily be shown that the Dirac-Hestenes equation can be written in
the form of Eq.(\ref{gm}), where the \textquotedblleft
electric\textquotedblright\ and \textquotedblleft magnetic\textquotedblright%
\ like currents are non linear functionals depending on $S,F$ and $P$
\cite{rod,rodoliv2007}.

A last observation is necessary. We are sure that an attentive reader which
has not been yet introduced to the Clifford bundle formalism may have become
intrigued with our statement in the abstract that pair forms may be used
coherently besides in electromagnetism, also in any other physical theory.
\ We just mentioned that Dirac equation can be represented \ by sum of
nonhomogeneous even sections of the Clifford bundle. But, someone may
still\ eventually ask: and what about Einstein's gravitational theory which is
formulated with a \textit{symmetric }metric field as its fundamental field?
Well, gravitational theory may also be formulated \ that field being
represented by a set of four linearly independent \ $1$-form fields living on
$M$

The details of how this is done can be found in, e.g.,
\cite{nr,qr,rodoliv2007} and that fact seems to give even more importance to
the modern theory of differential forms which started with Cartan.

\appendix

\section{How to Calculate $F$}

\subsection{Green's Identity for Differential Forms}

In this section $M$ is a $n$-dimensional differentiable manifold and
$\mathbf{g\in}\sec T_{0}^{2}M$ is a metric on $M$ of arbitrary signature
$(p,q)$, with $p+q=n$. Moreover, we denote by \texttt{g}$\in\sec T_{2}^{0}M$
the metric on the cotangent bundle such that \ in an arbitrary coordinate
basis where $\mathbf{g=}g_{\mu\nu}dx^{\mu}\otimes dx^{\nu}$ and \texttt{g
}$=g^{\mu\nu}\frac{\partial}{\partial x^{\mu}}\otimes\frac{\partial}{\partial
x^{\nu}}$, it is $g_{\mu\nu}g^{\mu\alpha}=\delta_{\nu}^{\alpha}$. We suppose
moreover that $\bigwedge T^{\ast}M$ and ${\mathcal{C}\!\ell}(M,\mathtt{g})$
are respectively the exterior and Clifford algebra bundles of $M$. Let
$P\in\sec\bigwedge^{p}T^{\ast}M\subset\sec{\mathcal{C}\!\ell}(M,\mathtt{g})$.
We shall derive an integral identity involving $P$, $dP$ a $\delta P$ and a
Green (\emph{extensor}) distribution\footnote{The distribution $\mathbf{G}%
_{\breve{x}}$ is also called a $p$-form-valued de Rham current. Rigorously we
should write $P\in\sec\mathfrak{D}^{\prime}(M,\bigwedge\nolimits^{p}T^{\ast
}M)\subset\sec\mathfrak{D}^{\prime}(M,{\mathcal{C}\!\ell}(M,\mathtt{g}))$ and
$\mathbf{G}_{\breve{x}}\in\sec\bigwedge\nolimits^{p}T^{\ast}\breve{M}%
\otimes\sec\mathfrak{D}^{\prime}(M,\bigwedge\nolimits^{n-p}T^{\ast
}M)\hookrightarrow\sec\bigwedge\nolimits^{p}T^{\ast}\bar{M}\otimes\sec
D^{\prime}(M,C\ell(M,\mathtt{g}))$ where $\sec\mathfrak{D}^{\prime
}(M,\bigwedge\nolimits^{n-p}T^{\ast}M)$ is the space of the linear functionals
over the sections of $\bigwedge\nolimits^{p}T^{\ast}M$ of $p$-forms of compact
support (in the sense of its action as, e.g., in Eq.(\ref{3nn.1}). $\breve{M}$
is a copy of $M$ and is there to recall that $\mathbf{G}_{\breve{x}}$ is a two
point distribution.} $\mathbf{G}_{\breve{x}}\in\sec\bigwedge\nolimits^{p}%
T^{\ast}\breve{M}\otimes\sec\bigwedge\nolimits^{n-p}T^{\ast}M$ that is a
generalization of the well known Green's identities of classical vector
calculus. This identity is crucial in order to obtain a formula solving
certain differential equations satisfied by $P$.

Let $\{\theta^{j},\theta_{j}\}$ be a pair reciprocal bases for $\bigwedge
^{1}T^{\ast}M\hookrightarrow{\mathcal{C}\!\ell}(M,\mathtt{g})$. In what
follows the notation $\breve{\theta}_{i}$ ($\breve{\theta}^{i}$) means that
these forms are calculated at a point $\breve{x}\in\breve{M}$. Now, we
introduce the Dirac extensor distribution $\mathbf{\delta}_{\breve{x}}\in
\sec\bigwedge\nolimits^{p}T^{\ast}\breve{M}\otimes\sec\bigwedge\nolimits^{n-p}%
T^{\ast}M$ by
\begin{equation}
\int\mathbf{\delta}_{\breve{x}}\wedge P(x)=P(\breve{x}). \label{3nn.1}%
\end{equation}
where $\mathbf{\delta}_{\breve{x}}$ has support only at $\breve{x}$. If
$\{x^{i}\}$ are the coordinate of a chart of an atlas of $M$ and if we choose
$\{\theta^{j}=dx^{j},\theta_{j}=g_{ij}dx^{i}\}$ then we can easily verify
that
\begin{align}
\mathbf{\delta}_{\breve{x}}  &  =\frac{(-1)^{p(n-p)}}{p!}\breve{\theta}%
_{i_{1}\ldots i_{p}}\otimes\star\theta^{i_{1}\ldots i_{p}}\delta(x-\breve
{x}),\nonumber\\
\mathbf{\delta}(x-\bar{x})  &  =\mathbf{\delta}(x^{1}-\breve{x}^{1}%
)\ldots\mathbf{\delta}(x^{n}-\breve{x}^{n}),\nonumber\\
\breve{\theta}_{i_{1}\ldots i_{p}}  &  =\breve{\theta}_{i_{1}}\wedge
\cdots\wedge\breve{\theta}_{i_{p}},\text{ }\theta^{i_{1}\ldots i_{p}}%
=\theta^{i_{1}}\wedge\cdots\wedge\theta^{i_{p}}, \label{3nn.2}%
\end{align}
where in Eq.(\ref{3nn.2}) $\mathbf{\delta}(x^{i}-\breve{x}^{i}),$
$i=1,2,\ldots, n$ are the usual (scalar) Dirac measures.

The Green distribution is supposed to satisfy the following differential equation%

\begin{equation}
\mathbf{\square G}_{\breve{x}}=-(d\delta+\delta d)\mathbf{G}_{\breve{x}%
}=\mathbf{\delta}_{\breve{x}}.
\end{equation}

We now prove the following identity:
\begin{align}
\mathbf{\delta}_{\breve{x}}\wedge P  &  =(-1)^{n+p}[d\mathbf{G}_{\breve{x}%
}\wedge\delta P-\delta\mathbf{G}_{\breve{x}}\wedge dP]\nonumber\\
&  \hspace{-0.6cm}-d[\delta\mathbf{G}_{\breve{x}}\wedge P-(-1)^{np+p+s+1}%
\underset{\tau_{\mathbf{%
%TCIMACRO{\TeXButton{g}{\sslg}}%
%BeginExpansion
\sslg
%EndExpansion
}}}{\star}P\wedge\underset{\tau_{\mathbf{%
%TCIMACRO{\TeXButton{g}{\sslg}}%
%BeginExpansion
\sslg
%EndExpansion
}}}{\star}d\mathbf{G}_{\breve{x}}]. \label{3nn.5}%
\end{align}

We start with the product $d\mathbf{G}_{\breve{x}}\wedge\delta P$ and make
some transformations on it using the definition of the Hodge coderivative and
some other well known identities involving the exterior product\footnote{See,
e.g., Section 2.4.2 of \cite{rodoliv2007}.}. We then have
\begin{align*}
d\mathbf{G}_{\breve{x}}\wedge\delta P  &  =(-1)^{n(p+1)+s+1}d\mathbf{G}%
_{\breve{x}}\wedge\underset{\tau_{\mathbf{%
%TCIMACRO{\TeXButton{g}{\sslg}}%
%BeginExpansion
\sslg
%EndExpansion
}}}{\star}d\underset{\tau_{\mathbf{%
%TCIMACRO{\TeXButton{g}{\sslg}}%
%BeginExpansion
\sslg
%EndExpansion
}}}{\star}P\\
&  =(-1)^{np+n+s+1}d\underset{\tau_{\mathbf{%
%TCIMACRO{\TeXButton{g}{\sslg}}%
%BeginExpansion
\sslg
%EndExpansion
}}}{\star}P\wedge\underset{\tau_{\mathbf{%
%TCIMACRO{\TeXButton{g}{\sslg}}%
%BeginExpansion
\sslg
%EndExpansion
}}}{\star}d\mathbf{G}_{\bar{x}}\\
&  =(-1)^{s+1}d(\underset{\tau_{\mathbf{%
%TCIMACRO{\TeXButton{g}{\sslg}}%
%BeginExpansion
\sslg
%EndExpansion
}}}{\star}P\wedge\underset{\tau_{\mathbf{%
%TCIMACRO{\TeXButton{g}{\sslg}}%
%BeginExpansion
\sslg
%EndExpansion
}}}{\star}d\mathbf{G}_{\breve{x}})-(-1)^{n+p}\delta d\mathbf{G}_{\breve{x}%
}\wedge P\\
&  =(-1)^{s+1}d(\underset{\tau_{\mathbf{%
%TCIMACRO{\TeXButton{g}{\sslg}}%
%BeginExpansion
\sslg
%EndExpansion
}}}{\star}P\wedge\underset{\tau_{\mathbf{%
%TCIMACRO{\TeXButton{g}{\sslg}}%
%BeginExpansion
\sslg
%EndExpansion
}}}{\star}d\mathbf{G}_{\breve{x}})+(-1)^{n+p}[(-\delta d-d\delta
)\mathbf{G}_{\breve{x}}\wedge P]\\
&  +(-1)^{n+p}d\delta\mathbf{G}_{\breve{x}}\wedge P\\
&  =(-1)^{s+1}d(\underset{\tau_{\mathbf{%
%TCIMACRO{\TeXButton{g}{\sslg}}%
%BeginExpansion
\sslg
%EndExpansion
}}}{\star}P\wedge\underset{\tau_{\mathbf{%
%TCIMACRO{\TeXButton{g}{\sslg}}%
%BeginExpansion
\sslg
%EndExpansion
}}}{\star}d\mathbf{G}_{\breve{x}})+(-1)^{n+p}\delta_{\breve{x}}\wedge
P+(-1)^{n+p}d(\delta\mathbf{G}_{\breve{x}}\wedge P)\\
&  +\delta\mathbf{G}_{\breve{x}}\wedge dP,
\end{align*}
from where Eq.(\ref{3nn.5}) follows.

Integrating both sides on the $n$-dimensional region $\mathcal{U}\subset M$ we
have\footnote{Analogous equation to Eq.(\ref{3nn.6}) appears in Thirring's
book \cite{thirring}. However take care on comparing the equations there and
here because of some $(-1)$ signs arising due to different definitions of the
Hodge coderivative.}%

\begin{align}
P(\breve{x})  &  =(-1)^{n+p}\int\nolimits_{\mathcal{U}}[d\mathbf{G}_{\breve
{x}}\wedge\delta P-\delta\mathbf{G}_{\breve{x}}\wedge dP]\nonumber\\
&  -\int\nolimits_{\partial\mathcal{U}}\delta\mathbf{G}_{\breve{x}}\wedge
P-(-1)^{n+p+s}\underset{\tau_{\mathbf{%
%TCIMACRO{\TeXButton{g}{\sslg}}%
%BeginExpansion
\sslg
%EndExpansion
}}}{\star}d\mathbf{G}_{\breve{x}}\wedge\underset{\tau_{\mathbf{%
%TCIMACRO{\TeXButton{g}{\sslg}}%
%BeginExpansion
\sslg
%EndExpansion
}}}{\star}P]. \label{3nn.6}%
\end{align}

\subsection{Solution of ${%
%TCIMACRO{\TeXButton{d}{\mbox{\boldmath$\partial$}}}%
%BeginExpansion
\mbox{\boldmath$\partial$}%
%EndExpansion
}F=J$}

We now applies the above formula for solving the equation ${%
%TCIMACRO{\TeXButton{d}{\mbox{\boldmath$\partial$}}}%
%BeginExpansion
\mbox{\boldmath$\partial$}%
%EndExpansion
}F=J$ in a Minkowski manifold. We start by choosing a chart with coordinates
$\{\mathrm{x}^{\mu}\}$ in the \textit{ELPG. We write as in the text }%
$\gamma^{\mu}=d\mathrm{x}^{\mu}$ and $\gamma_{\mu}=\eta_{\mu\nu}\gamma^{\nu}$.
Then, since ${%
%TCIMACRO{\TeXButton{d}{\mbox{\boldmath$\partial$}}}%
%BeginExpansion
\mbox{\boldmath$\partial$}%
%EndExpansion
}F=J$ is equivalent to $dF=0$ and $\delta F=-J$, we have using the retarded
solution
\begin{align}
\mathbf{G}_{s}(\mathrm{x-\breve{x}})  &  =\frac{1}{2}\breve{\gamma}_{\mu
_{1}\mu_{2}}\otimes\underset{\tau_{\mathbf{%
%TCIMACRO{\TeXButton{g}{\sslg}}%
%BeginExpansion
\sslg
%EndExpansion
}}}{\star}\gamma^{\mu_{1}\mu_{2}}G_{s}(\mathrm{x-\breve{x}}),\nonumber\\
{%
%TCIMACRO{\TeXButton{d}{\mbox{\boldmath$\partial$}}}%
%BeginExpansion
\mbox{\boldmath$\partial$}%
%EndExpansion
}^{2}G_{s}(\mathrm{x-\breve{x}})  &  =\delta(\mathrm{x-\breve{x}})
\end{align}
where $G_{s}$ is the scalar Green \emph{retarded} function (see,
e.g.,\cite{vla}), which vanishes outside the light cone at $x$. Then%

\begin{equation}
F(\mathrm{x})=-\int\nolimits_{\mathcal{U}}d\mathbf{G}_{s}(\mathrm{x-\breve{x}%
})\wedge J(\mathrm{\breve{x})}-\int\limits_{\partial\mathcal{U}}%
\delta\mathbf{G}_{s}(\mathrm{x-\breve{x}})\wedge F(\mathrm{\breve{x}})+\text{
}\underset{\tau_{\mathbf{%
%TCIMACRO{\TeXButton{g}{\sslg}}%
%BeginExpansion
\sslg
%EndExpansion
}}}{\star}d\mathbf{G}_{s}(\mathrm{x-\breve{x}})\wedge\underset{\tau_{\mathbf{%
%TCIMACRO{\TeXButton{g}{\sslg}}%
%BeginExpansion
\sslg
%EndExpansion
}}}{\star}F(\mathrm{\breve{x}}), \label{8n5}%
\end{equation}
and supposing that $F$ vanishes on the boundary $\partial\mathcal{U}$ we end
with \textit{ }%
\begin{equation}
F(\mathrm{x})=-\int\nolimits_{\mathcal{U}}d\mathbf{G}(\mathrm{x-\breve{x}%
})\wedge J(\mathrm{\breve{x}).} \label{F}%
\end{equation}
This equation shows explicitly that $F\rightarrow-F$ when we decide to
\textit{relabel} the charges entering $J$ from $q^{(i)}$ to $-q^{(i)}$
something that as already discussed in the text happens if we calculate $%
%TCIMACRO{\dint }%
%BeginExpansion
{\displaystyle\int}
%EndExpansion
\underset{\tau_{\mathbf{%
%TCIMACRO{\TeXButton{g}{\sslg}}%
%BeginExpansion
\sslg
%EndExpansion
}}}{\star}J$ in a chart with a different orientation than the positive one
defined by $\gamma^{0}\wedge\gamma^{1}\wedge\gamma^{2}\wedge\gamma^{3}$.

\end{document}